\documentclass[twocolumn]{aastex63}
\usepackage{amsmath,amssymb}
\usepackage{hyperref}
\usepackage{gensymb}
\received{XXX}
\revised{XXX}
\accepted{\today}
\submitjournal{ApJ}

\shorttitle{Low states in IPs}
\shortauthors{Covington et al.}
\graphicspath{{./}{Figures/}}

\begin{document}

\title{Investigating the low-flux states in six Intermediate Polars}

\correspondingauthor{A. W. Shaw}
\email{aarrans@unr.edu}

\author{Ava E. Covington}
\affiliation{Department of Physics, University of Nevada, Reno, NV 89557, USA}

\author[0000-0002-8808-520X]{Aarran W. Shaw}
\affiliation{Department of Physics, University of Nevada, Reno, NV 89557, USA}

\author[0000-0002-8286-8094]{Koji Mukai}
\affiliation{CRESST and X-ray Astrophysics Laboratory, NASA Goddard Space Flight Center, Greenbelt, MD 20771, USA}
\affiliation{Department of Physics, University of Maryland, Baltimore County, 1000 Hilltop Circle, Baltimore, MD 21250, USA}

\author[0000-0001-7746-5795]{Colin Littlefield}
\affiliation{Department of Physics, University of Notre Dame, Notre Dame, IN 46556, USA}
\affiliation{Department of Astronomy, University of Washington, Seattle, WA 98195, USA}
\affiliation{Bay Area Environmental Research Institute, Moffett Field, CA 94035 USA}

\author[0000-0003-3944-6109]{Craig O. Heinke}
\affiliation{Department of Physics, University of Alberta, Edmonton, AB T6G 2G7, Canada}

\author[0000-0002-7092-0326]{Richard M. Plotkin}
\affiliation{Department of Physics, University of Nevada, Reno, NV 89557, USA}

\author{Doug Barrett}
\affiliation{American Association of Variable Star Observers, 49 Bay State Road, Cambridge, MA 02138, USA}
\affiliation{Observatoire Marouzeau, 23300 Creuse, France}

\author{James Boardman}
\affiliation{American Association of Variable Star Observers, 49 Bay State Road, Cambridge, MA 02138, USA}
\affiliation{Luckydog Observatory, 65012 Howath Rd, De Soto, WI
54624, USA}

\author{David Boyd}
\affiliation{American Association of Variable Star Observers, 49 Bay State Road, Cambridge, MA 02138, USA}
\affiliation{British Astronomical Association, Variable Star Section, West Challow Observatory, OX12 9TX, UK}

\author{Stephen M. Brincat}
\affiliation{American Association of Variable Star Observers, 49 Bay State Road, Cambridge, MA 02138, USA}
\affiliation{Flarestar Observatory (MPC 171), Fl. 5/B, George Tayar Street, San Gwann, SGN 3160, Malta}

\author{Rolf Carstens}
\affiliation{American Association of Variable Star Observers, 49 Bay State Road, Cambridge, MA 02138, USA}
\affiliation{Geyserland Observatory E89, 120 Homedale Street, Pukehangi, Rotorua 3015, New Zealand}
\affiliation{Royal Astronomical Society of New Zealand}

\author{Donald F. Collins}
\affiliation{American Association of Variable Star Observers, 49 Bay State Road, Cambridge, MA 02138, USA}
\affiliation{Warren Wilson College, 701 Warren Wilson Road, Swannanoa, NC 28778}

\author{Lewis M. Cook}
\affiliation{American Association of Variable Star Observers, 49 Bay State Road, Cambridge, MA 02138, USA}

\author{Walter R. Cooney}
\affiliation{American Association of Variable Star Observers, 49 Bay State Road, Cambridge, MA 02138, USA}
\affiliation{Starry Night Observatory, Columbus, TX, USA}

\author{David Cejudo Fern\'{a}ndez}
\affiliation{American Association of Variable Star Observers, 49 Bay State Road, Cambridge, MA 02138, USA}
\affiliation{Observatorio El Gallinero, El Berrueco, Madrid, Spain}

\author{Sjoerd Dufoer}
\affiliation{American Association of Variable Star Observers, 49 Bay State Road, Cambridge, MA 02138, USA}
\affiliation{Vereniging Voor Sterrenkunde (VVS), Oostmeers 122 C, 8000 Brugge, Belgium}

\author{Shawn Dvorak}
\affiliation{American Association of Variable Star Observers, 49 Bay State Road, Cambridge, MA 02138, USA}
\affiliation{Rolling Hills Observatory, Clermont, FL, USA}

\author{Charles Galdies}
\affiliation{American Association of Variable Star Observers, 49 Bay State Road, Cambridge, MA 02138, USA}
\affiliation{Institute of Earth Systems, University of Malta, Msida, Malta}

\author{William Goff}
\affiliation{American Association of Variable Star Observers, 49 Bay State Road, Cambridge, MA 02138, USA}

\author{Franz-Josef Hambsch}
\affiliation{American Association of Variable Star Observers, 49 Bay State Road, Cambridge, MA 02138, USA}
\affiliation{Vereniging Voor Sterrenkunde (VVS), Oostmeers 122 C, 8000 Brugge, Belgium}
\affiliation{Bundesdeutsche Arbeitsgemeinschaft f\"{u}r Ver\"{a}nderliche Sterne (BAV), Munsterdamm 90, 12169 Berlin, Germany}
\affiliation{Groupe Europ\'{e}en d'Observations Stellaires (GEOS), 23 Parc de Levesville, 28300 Bailleau l'Ev\^{e}que, France}

\author{Steve Johnston}
\affiliation{American Association of Variable Star Observers, 49 Bay State Road, Cambridge, MA 02138, USA}
\affiliation{British Astronomical Association, Variable Star Section}

\author{Jim Jones}
\affiliation{American Association of Variable Star Observers, 49 Bay State Road, Cambridge, MA 02138, USA}
\affiliation{CBA–Oregon, Jack Jones Observatory, 22665 Bents Road NE, Aurora, OR 97002, USA}

\author{Kenneth Menzies}
\affiliation{American Association of Variable Star Observers, 49 Bay State Road, Cambridge, MA 02138, USA}
\affiliation{Tigh Speuran Observatory, Framingham, MA 01701, USA}

\author{Libert A. G. Monard}
\affiliation{American Association of Variable Star Observers, 49 Bay State Road, Cambridge, MA 02138, USA}
\affiliation{Bronberg Observatory, CBA Pretoria, PO Box 11426, Tiegerpoort 0056, South Africa}
\affiliation{Kleinkaroo Observatory, Calitzdorp, St. Helena 1B, P.O. Box 281, 6660 Calitzdorp, Western Cape, South Africa}

\author{Etienne Morelle}
\affiliation{American Association of Variable Star Observers, 49 Bay State Road, Cambridge, MA 02138, USA}
\affiliation{Observatoire Sirene, 84400 Lagarde d'Apt, France}

\author{Peter Nelson}
\affiliation{American Association of Variable Star Observers, 49 Bay State Road, Cambridge, MA 02138, USA}
\affiliation{Ellinbank Observatory, Victoria, Australia.}

\author{Yenal \"{O}\u{g}men}
\affiliation{American Association of Variable Star Observers, 49 Bay State Road, Cambridge, MA 02138, USA}
\affiliation{Green Island Observatory, Ge\c{c}itkale Ma\u{g}usa, North Cyprus}

\author{John W. Rock}
\affiliation{American Association of Variable Star Observers, 49 Bay State Road, Cambridge, MA 02138, USA}
\affiliation{CBA Highworth Observatory, Highworth, Wiltshire, SN6 7PJ, UK}

\author{Richard Sabo}
\affiliation{American Association of Variable Star Observers, 49 Bay State Road, Cambridge, MA 02138, USA}
\affiliation{1344 Post Dr., Bozeman, MT 59715, USA}

\author{Jim Seargeant}
\affiliation{American Association of Variable Star Observers, 49 Bay State Road, Cambridge, MA 02138, USA}
\affiliation{The Albuquerque Astronomy Society (TAAS), P.O. Box 50581 Albuquerque, NM 87181, USA}

\author{Geoffrey Stone}
\affiliation{American Association of Variable Star Observers, 49 Bay State Road, Cambridge, MA 02138, USA}
\affiliation{First Light Observatory Systems, Corvallis, OR, USA}

\author{Joseph Ulowetz}
\affiliation{American Association of Variable Star Observers, 49 Bay State Road, Cambridge, MA 02138, USA}
\affiliation{Center for Backyard Astrophysics, 855 Fair Ln, Northbrook, IL 60062, USA}

\author{Tonny Vanmunster}
\affiliation{American Association of Variable Star Observers, 49 Bay State Road, Cambridge, MA 02138, USA}
\affiliation{CBA Belgium Observatory, Walhostraat 1A, B-3401 Landen, Belgium}
\affiliation{CBA Extremadura Observatory, E-06340 Fregenal de la Sierra, Badajoz, Spain}




\begin{abstract}

We present optical photometry of six intermediate polars that exhibit transitions to a low-flux state. For four of these systems, DW\,Cnc, V515\,And, V1223\,Sgr and RX\,J2133.7$+$5107, we are able to perform timing analysis in and out of the low states. We find that, for DW\,Cnc and V515\,And, the dominant periodicities in the light curves change as the flux decreases, indicating a change in the sources' accretion properties as they transition to the low state. For V1223\,Sgr we find that the variability is almost completely quenched at the lowest flux, but do not find evidence for a changing accretion geometry. For RX\,J2133.7$+$5107, the temporal properties do not change in the low state, but we do see a period of enhanced accretion that is coincident with increased variability on the beat frequency, which we do not associate with a change in the accretion mechanisms in the system. 

\end{abstract}



\section{Introduction} 
\label{sec:intro}

Intermediate polars (IPs) are a subclass of cataclysmic variables (CVs) in which a white dwarf (WD) accretes matter from a main sequence companion. In IPs the magnetic field of the WD is strong enough ($B\gtrsim10^6$ G) to disrupt the innermost regions of the accretion disk such that, at the magnetospheric boundary, matter flows along the magnetic field lines on to the poles of the WD \citep[see e.g.][for a review]{Patterson-1994}. Unlike polars, another class of magnetic CV (mCV), the spin period of the WD ($P_{\rm spin}$) and binary orbital period ($P_{\rm orb}$) are not synchronized. As a result, the power spectra of IP light curves are typically dominated by the WD spin frequency ($\omega=1/P_{\rm spin}$), the orbital frequency ($\Omega=1/P_{\rm orb}$) and the beat frequency ($\omega - \Omega$) and their respective harmonics.

Timing analysis of IP light curves, both at optical and X-ray wavelengths, can reveal a lot about the dominant mode of accretion in the system. Typically, IPs have a residual disk whose inner edge is truncated at the magnetospheric radius, and matter is forced along the field lines in so-called `accretion curtains.' This is the `disk-fed' model, and applies to almost all IPs. In the `stream-fed' model \citep{Hameury-1986}, on the other hand, there is no viscous accretion disk, and matter flows directly from the companion on to the WD magnetosphere. The diskless IP V2400\,Oph is often cited as the prototypical stream-fed system \citep{Hellier-2002a}. A third accretion mode, the `disk-overflow' model, is a hybrid of the two, where there is simultaneous stream- and disk-fed accretion on to the WD \citep{Lubow-1989,Hellier-1993,Armitage-1996,Armitage-1998}. 

\citet{Wynn-1992} and \citet{Ferrario-1999} both showed that the strengths of the different peaks in the power spectra of IP light curves (both optical and X-ray) could be used to determine the dominant accretion mode in the system. A typical, disk-fed IP will show strong variability on the spin period of the WD, thus the strongest peak in the (optical and X-ray) power spectrum will be at $\omega$. This is because, in the disk-fed mode, material is being forced on to field lines at a nearly uniform rate from a Keplerian disk. However, in the stream-fed model, the power spectra start to become dominated by multiples of the beat between the spin and orbital frequencies, due to the matter flowing directly from a fixed point in the binary rest frame rather than in the frame of the rotating WD. \citet{Ferrario-1999} show that, for diskless IPs, the X-ray light curves will show variability at $\omega$, $\omega-\Omega$ and $2\omega-\Omega$, with the strength of the peaks varying with viewing angle. The optical power spectra, on the other hand shows strong peaks at $\omega$, $\omega-2\Omega$ and $2(\omega-\Omega)$, with the power spectra of a lower inclination ($i\sim20^{\circ}$) system almost completely dominated by the $2(\omega-\Omega)$ peak.

Recently, observations of the IP FO\,Aqr have shown the system to fade by up to 2 magnitudes in the $V$-band, with comparable decreases in flux at X-ray energies \citep{Kennedy-2017,Littlefield-2020}. These so-called `low states' are relatively uncommon in IPs \citep[see e.g.][]{Garnavich-1988} compared to polars and are thought to be due to a temporary reduction in mass transfer from the donor \citep[potentially from starspots transiting the $L_1$ point on the surface of the donor;][]{Livio-1994}. FO\,Aqr has shown three such low states since 2015, offering the opportunity to study one in real time. \citet{Kennedy-2017} and \citet{Littlefield-2020} found that the timing signatures, at both optical and X-ray wavelengths, changed as the source progressed through its low states, indicative of a change in the dominant accretion mechanism from exclusively disk-fed to disk-overflow accretion as the flux decreased. Optical spectroscopy of the 2016 and 2017 low states confirmed this switch in accretion mode \citep{Kennedy-2020}. 

However, FO\,Aqr is only one of $\sim$50 confirmed IPs\footnote{according to the 2014 version of the IP catalog \href{https://asd.gsfc.nasa.gov/Koji.Mukai/iphome/catalog/omegaomega.html}{https://asd.gsfc.nasa.gov/Koji.Mukai/iphome/catalog\\/omegaomega.html}} and more IPs are starting to be observed in low states. For example, DW\,Cnc exhibited a low state that lasted $\sim$2 years, during which time optical spectroscopy revealed that the spin variability had dramatically weakened \citep{SeguraMontero-2020}. In addition, the IP DO\,Dra was not detected in a 55.4ks observation with the Nuclear Spectroscopic Telescope Array \citep[ NuSTAR;][]{Harrison-2013} in July 2018 \citep{Shaw-2020b}, a low state that was also detected at optical wavelengths \citep{Andronov-2018}. However, in the case of DO\,Dra, no coherent oscillations were detected. We also note that the IP TX\,Col has shown transitions to states of {\em enhanced} accretion \citep{Rawat-2021,Littlefield-2021}, during which time strong quasi-periodic oscillations (QPOs) dominated the optical light curves. These QPOs have been interpreted as due to a torus of diamagnetic blobs circling the (usually disk-overflow accreting) WD in unstable orbits \citep{Littlefield-2021}.

It is becoming evident that low states are more common in IPs than previously anticipated, and it has been shown that variability studies hold the key to understanding the accretion processes and how they may change. Therefore, in this work we present optical photometry of several IPs that exhibit low states and, where possible, present timing analysis in order to investigate the accretion processes at work.

\section{Observations and Analysis}
\label{sec:obs}

We utilized the public photometry database of the All-Sky Automated Survey for Supernovae \citep[ASAS-SN;][]{Shappee-2014,Kochanek-2017} and downloaded long-term light $V$- and $g$-band light curves of the IPs that have a 2--10 keV X-ray flux $F_{\rm X}\gtrsim1\times10^{-11}$ erg cm$^{-2}$ s$^{-1}$ in their `normal' state.\footnote{\href{https://asd.gsfc.nasa.gov/Koji.Mukai/iphome/catalog/omegaomega.html}{https://asd.gsfc.nasa.gov/Koji.Mukai/iphome/catalog\\/omegaomega.html}} Such a sample can be assumed to be representative of the intrinsic IP population \citep{Pretorius-2014}. We checked for potential low states (which we define as a sustained drop in flux of $\gtrsim0.5$ mag from the average) in each light curve. Upon finding evidence for a low state in a source, we then checked the photometry database of the American Association of Variable Star Observers (AAVSO) for higher cadence data with which we could perform timing analysis. The AAVSO data also often went back further than ASAS-SN, opening up the possibility of probing multiple low states (see e.g. V515\,And). Additionally, AAVSO light curves often allowed us to probe higher frequencies than allowed by ASAS-SN, which has a typical cadence of $\sim1$d for many objects, thus allowing us to study variability on the spin and beat frequencies. The majority of AAVSO data presented in this work is photometry in the $CV$-band, i.e. unfiltered data with a $V$-band zeropoint. However, we do include some $V$-, $R$- and $CR$-band (unfiltered with an $R$-band zeropoint) photometry too. The sources we focus on in this work are introduced in Section \ref{sec:sources}. Full details of the AAVSO observations we used in this work are presented in Table \ref{tab:obs}.

For one source, DO\,Dra (see Section \ref{sec:sources}), we also include photometry from the Neil Gehrels Swift Observatory Ultraviolet/Optical Telescope \citep[Swift/UVOT;][]{Roming-2005}. Swift/UVOT is a 30cm diameter telescope that operates simultaneously with the spacecraft's X-ray Telescope. There are 17 Swift observations of DO\,Dra in the archive for a total of 54 exposures split between the UV filters ($UVW2$,$UVM2$,$UVW1$,$U$). Aperture photometry was performed using {\tt uvotsource} as part of {\sc heasoft} v6.27.2. We extracted photons from a 5\arcsec\ radius circular region centered on the source and the background was measured using a 20\arcsec\ radius circular region centered on a source free region of the CCD. All reported magnitudes are in the Vega system and uncertainties include the statistical $1\sigma$ error and a systematic uncertainty that accounts for the uncertainty in the shape of the UVOT point spread function (PSF).

For sources with sufficient AAVSO data we split the data in to several observing epochs, typically separated by prolonged gaps between observing campaigns by AAVSO observers (usually due to visibility). To construct the AAVSO light curves we then filtered data in each epoch so that no data overlapped in time in the case of two or more observers observing a source simultaneously. Photometric data that overlap in time but have slight differences in calibration may manifest as false signals in the timing analysis on the time-scales that we are probing, so we filtered the data by observer for each light curve epoch. For each epoch we first prioritized the observer who provided the most data, then included non-simultaneous data from all other observers that observed the source during the same epoch.

We performed timing analysis on the AAVSO light curves to search for changes in the dominant periodicity across observing epochs. To search for these periodicities we use Lomb Scargle periodogram analysis \citep{Lomb-1976, Scargle-1982}, which utilizes least-squares fitting of sinusoids to the light curve to determine the power at each frequency in a given range. We implement this analysis using the {\tt LombScargle} class included in the {\sc astropy python} package \citep{Astropy-2013,Astropy-2018}.  
Before performing the Lomb Scargle analysis, we applied a heliocentric correction to the data using the {\tt time} package available in {\sc astropy}. As AAVSO data are from telescopes distributed around the world, we corrected all photometric data as if they were observed from the W. M. Keck observatory on Mauna Kea, USA. This is a reasonable correction to make, as the maximum time correction for two antipodean observers observing the same source at the same time (if that were possible) is only $\sim0.04$ s, and we are searching for periodicity differences on $>>10$ s timescales.

For each periodogram we calculated a 99.9\% significance threshold for the power. We did this by randomly shuffling the light curve magnitudes but keeping the time-stamps the same for each epoch, effectively creating a randomized light curve with the same sampling as the original data. We calculated the peak Lomb Scargle power for 10,000 of these randomized light curves, again for every epoch of AAVSO data, from which we derived the 99.9\% significance. Lomb Scargle periodograms are presented for four of the six sources studied in this work.

\begin{table*}[]
    \centering
    \begin{tabular}{lcll}
    \hline
    {\bf Source} & {\bf MJD range} & {\bf Bands} & {\bf AAVSO observer codes} \\
    \hline\hline
         DW\,Cnc & 57044 -- 59235 & $CV$  &  {\scriptsize CDZ, COO, CWT, DFS, DKS, GFB, JJI,}\\
         & & & {\scriptsize MGW, MZK, RJWB, SRIC, UJHA, VMT} \\
         \hline
         V515\,And & 55464 -- 59251 & $CV$  & {\scriptsize BDG, BJAA, BPO, BSM, CDK, CDZ, COO,} \\
         & & & {\scriptsize CWT, DFS, DKS, GCHB, JJI, MEV, OYE,} \\
         & & & {\scriptsize RJWB, SGEA, SJIA, UJHA, VMT} \\
         \hline
         V1223\,Sgr & 56860 -- 59181 & $V$, $CV$ & {\scriptsize CROA, DKS, HMB, MGW, NLX} \\
         \hline
         RX\,J2133$+$5107 & 55460 -- 59333 & $CV$ & {\scriptsize BDG, BJAA, BPO, CDZ, COO, DFS, DKS,}\\
         & & & {\scriptsize GFB, JJI, MEV, RJWB, SRIC, VMT} \\
         \hline
         DO\,Dra & 55668 -- 59011 & $CV$, $V$, $R$ & {\scriptsize COO,  MEV, MNIC, JJI, JSJA, UJHA} \\
         \hline
         V1025\,Cen & 55211 -- 59165 & $CR$ & {\scriptsize MLF} \\
    \hline
    \end{tabular}
    \caption{Table of AAVSO observations. The AAVSO observer codes identify the individual observers whose data we used in this work.}
    \label{tab:obs}
\end{table*}

\subsection{Sources}
\label{sec:sources}
In this work we focus on 6 sources that showed at least one low state in ASAS-SN data since the project began in 2014. We choose not to include FO\,Aqr in this analysis as its low states have been well-studied previously \citep{Kennedy-2017,Littlefield-2020,Kennedy-2020}. 

DW\,Cnc is classified as a low luminosity IP \citep[LLIP;][]{Mukai-2017}, a subclass of IPs so named for their low X-ray luminosities ($L_{\rm X}\sim10^{31}$ erg s$^{-1}$) relative to most IPs. \citet{Patterson-2004} reported spectroscopic periods of 86.1 and 38.6 min, interpreted as $P_{\rm orb}$ and $P_{\rm spin}$, respectively \citep[see also][]{RodriguezGil-2004}, as well as a 69.9 min photometric period consistent with the beat period ($P_{\rm beat}$). As discussed in Section \ref{sec:intro}, DW\,Cnc underwent a transition to a low state in 2018, accompanied by a weakening of the WD spin signal in time resolved optical spectra. \citet{SeguraMontero-2020} hypothesize that DW\,Cnc is a magnetic analog of VY\,Scl systems - a class of CVs that exhibit occasional low states due to episodes of reduced mass transfer from the companion star. In this work we supplement the spectroscopic study of DW\,Cnc by \citet{SeguraMontero-2020} with timing analysis of the AAVSO $CV$-band photometry in and out of the low state.

V515\,And was identified as the optical counterpart to the X-ray source XSS\,J00564$+$4548, with strong oscillations on a period of $\sim480$ s \citep{Bikmaev-2006}. This was later refined to $P_{\rm spin}=465.5$ s and $P_{\rm beat}=488.6s$, inferring $P_{\rm orb}=2.73$ h \citep{Kozhevnikov-2012}. Prior to this work, no low states of V515\,And have been reported.

V1223\,Sgr is arguably one of the most well-studied IPs, it was identified as the $V\sim13$ optical counterpart to the {\em Uhuru} X-ray source 4U\,1849$-$41 \citep{Steiner-1981}. The WD spin variability is more prominent in the X-ray light curves \citep[$P_{\rm spin}=745.6$ s;][]{Osborne-1985} than in the optical. However, \citet{Jablonski-1987} note that $P_{\rm beat}=794.4$ is stronger in the optical band. V1223\,Sgr has a well-documented history of low states, having shown numerous excursions to low fluxes (fainter than its typical $B\sim12$) in photographic plate observations since 1931 \citep{Garnavich-1988,Simon-2014}. None of the low-state studies thus far have included timing analysis down to the level of $P_{\rm spin}$. However, \citet{Simon-2014} did find a typical recurrence period of $\sim1092$ d for low states during the period of observations spanning 1999-2009. In this work, we focus on the two most recent low states exhibited by V1223\,Sgr, one of which lasted $\gtrsim1$ year and showed a decrease in flux of $\sim2$ mag and the second of which was a much shorter duration, lower amplitude low state that occurred almost immediately after the source had recovered from the previous one.

RX\,J2133.7$+$5107 (1RXS\,J213344.1$+$510725) was discovered in the ROSAT Galactic Plane Survey as an optically bright CV \citep{Motch-1998}. It was confirmed as an IP with $P_{\rm spin}=570.8$ s and $P_{\rm orb}=7.14$ h \citep{Bonnet-Bidaud-2006,Thorstensen-2010}. Prior to this work, no low states have been reported for this system.

DO\,Dra (often also referred to as YY\,Dra in literature) is, like DW\,Cnc, classified as a LLIP \citep{Mukai-2017}, with $P_{\rm orb}=3.96$ h and $P_{\rm spin}=529.3$ s \citep{Haswell-1997}. However, unusually for an IP, only harmonics of the spin and beat are typically observed in the power spectra of DO\,Dra, and $P_{\rm spin}$ often goes unseen, at X-ray, optical and UV wavelengths \citep[see e.g.][]{Patterson-1993,Haswell-1997}. The source has been seen to transition to a low state \citep{Andronov-2018}, though no coherent oscillations were seen in the $V$-band.

The LLIP V1025\,Cen was identified as the optical counterpart to the X-ray source RX\,J1238$-$38, with $P_{\rm spin}=35.8$ min and $P_{\rm orb}=1.41$ h \citep{Buckley-1998,Hellier-1998}, making the source one of the few IPs below the period gap. \citet{Ferrario-1999} note that V1025\,Cen poses a problem to simple disk-fed and stream-fed models of accretion in IPs, as the system exhibits characteristics of both \citep[see also][]{Hellier-2002b}. Prior to this work, no low states of V1025\,Cen have been studied in detail\citep[though see][who probe the source at optical wavelengths during its recovery from the low state]{Littlefield-2022}.

\section{Results}
\label{sec:results}

\subsection{DW\,Cnc}
\label{sec:DWCnc}

\begin{figure*}
    \centering
    \includegraphics[width=\textwidth]{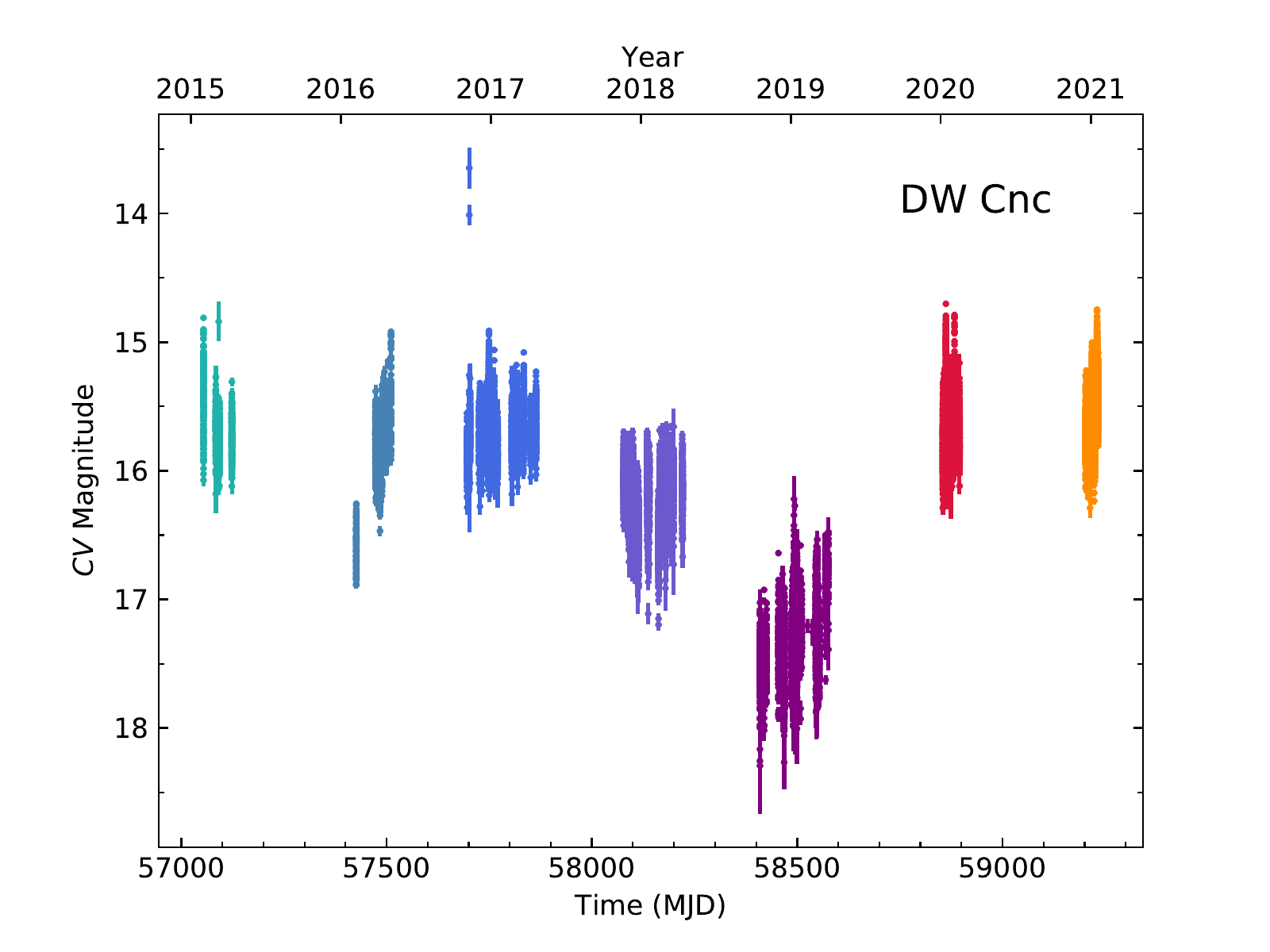}
    \caption{AAVSO light curve of DW\,Cnc from 2015--2021. Observations were taken in a clear filter and mapped on to the $V$-band ($CV$). The colors represent the different epochs used for timing analysis.}
    \label{fig:DWCnc_LC}
\end{figure*}

The AAVSO $CV$-band light curve of DW\,Cnc is shown in Fig. \ref{fig:DWCnc_LC}. This is similar to the light curve presented by \citet{SeguraMontero-2020}, but over a shorter time period. We focus on the 2015-2021 date range, where there are a high concentration of observations with well-constrained uncertainties. DW\,Cnc appears to begin to decline in flux around or before MJD$\sim$58080 (Fig. \ref{fig:DWCnc_LC}), before it reached its lowest flux ($CV\sim$17.5) around MJD $\sim58400$ and began to rise again. The source had recovered to its typical optical flux ($CV\sim$15.5) by MJD $\sim$58850. We also note the possibility of a short-lived transition to a lower-than-typical flux (though not as low as the 2018--2019) low state) in early 2016.

\begin{figure}
    \centering
    \includegraphics[width=0.5\textwidth]{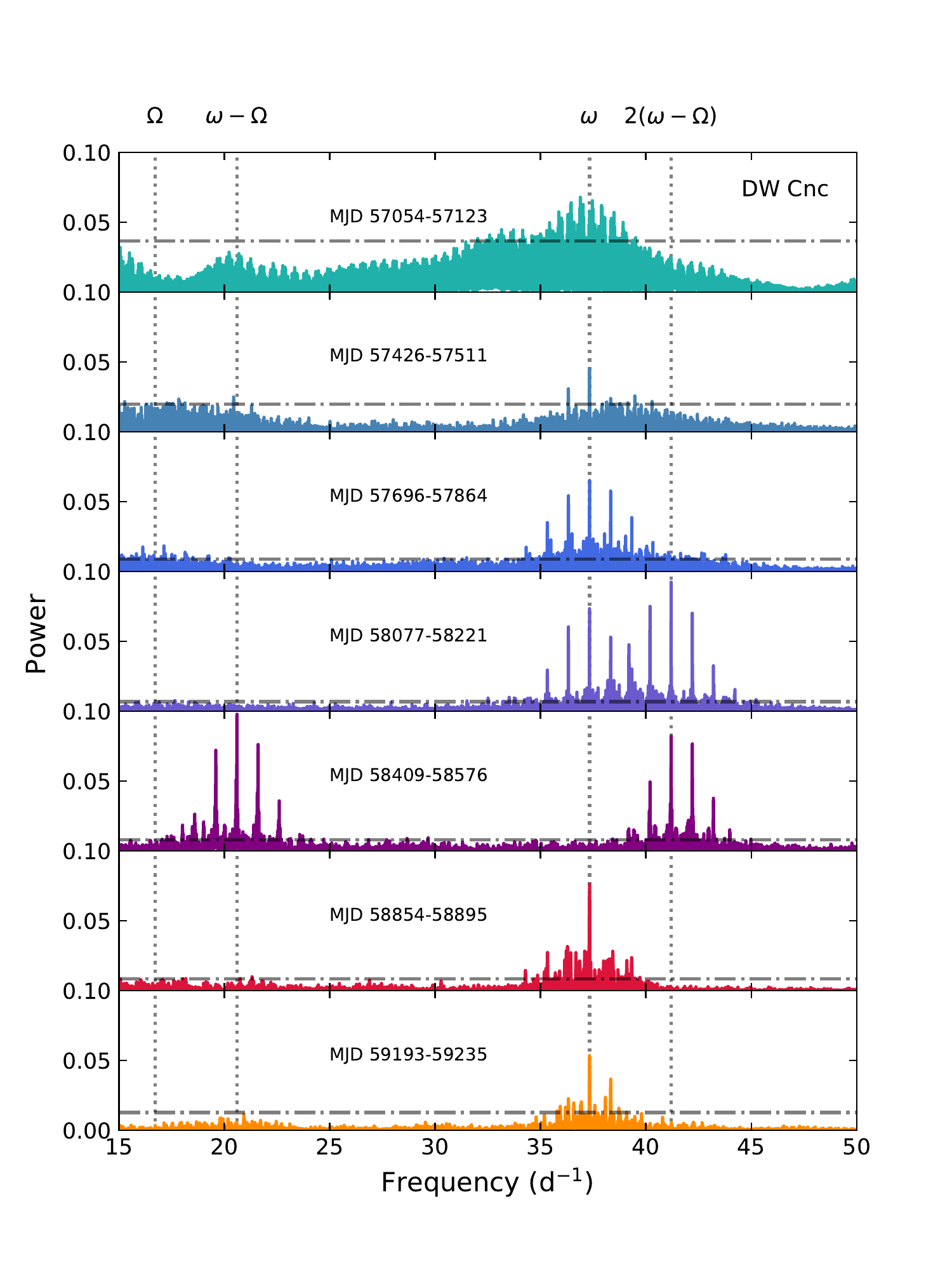}
    \caption{Lomb-Scargle periodograms of the $CV$-band light curves of DW\,Cnc. The colors match the epochs defined in Fig. \ref{fig:DWCnc_LC} and the relevant MJD ranges are labeled in each panel. The vertical dotted lines indicate known periodicities related to the system and the dot-dashed horizontal lines indicate the 99.9\% confidence level.}
    \label{fig:DWCnc_periodograms}
\end{figure}

Lomb Scargle periodograms of each colored epoch are presented in Fig. \ref{fig:DWCnc_periodograms}. Labeled are characteristic frequencies associated with the system, $\Omega=16.725$ d$^{-1}$ ($P_{\rm orb}=86.1$ min), $\omega=37.327$ d$^{-1}$ ($P_{\rm spin}=38.6$ min) and integer multiples of the beat frequency \citep{RodriguezGil-2004,Patterson-2004}. In the majority of the panels of Fig. \ref{fig:DWCnc_periodograms} we see that the light curves are dominated by $\omega$ (though in the top panel, the power at $\omega$ is severely aliased). However, as the source flux starts its decline to the low state in 2018 (fourth panel), we see strong variability emerge at $2(\omega-\Omega)$, before the signal from the spin of the WD completely disappears at the lowest flux (fifth panel) and the periodogram is dominated by power at $\omega-\Omega$ and $2(\omega-\Omega)$. By the time the flux recovers, the optical light curve of DW\,Cnc is once again dominated by $\omega$. 

Changes in the dominant periodicities in DW\,Cnc are also noted by \citet{SeguraMontero-2020}, who performed optical spectroscopy of the source during the decline to the low state (MJD 58308 and 58137--58143). They noted that the beat and spin periods were much weaker than typical. From analysing the photometry, we confirm the weakening, even disappearance, of the spin variability in the low state. However we find that the beat (and double the beat) frequency, does not weaken, in fact it becomes much stronger as the source flux declines. \citet{SeguraMontero-2020} predicted that DW\,Cnc would regain its IP characteristics (i.e. show strong variability on $\omega$) after the low state had finished. We show here that this prediction was indeed fulfilled in the final two epochs of AAVSO photometry. \citet{Duffy-2021} also report the return of the spin signal and also note three short-lived outbursts that are not covered by the AAVSO observations.

\subsection{V515\,And}

\begin{figure*}
    \centering
    \includegraphics[width=\textwidth]{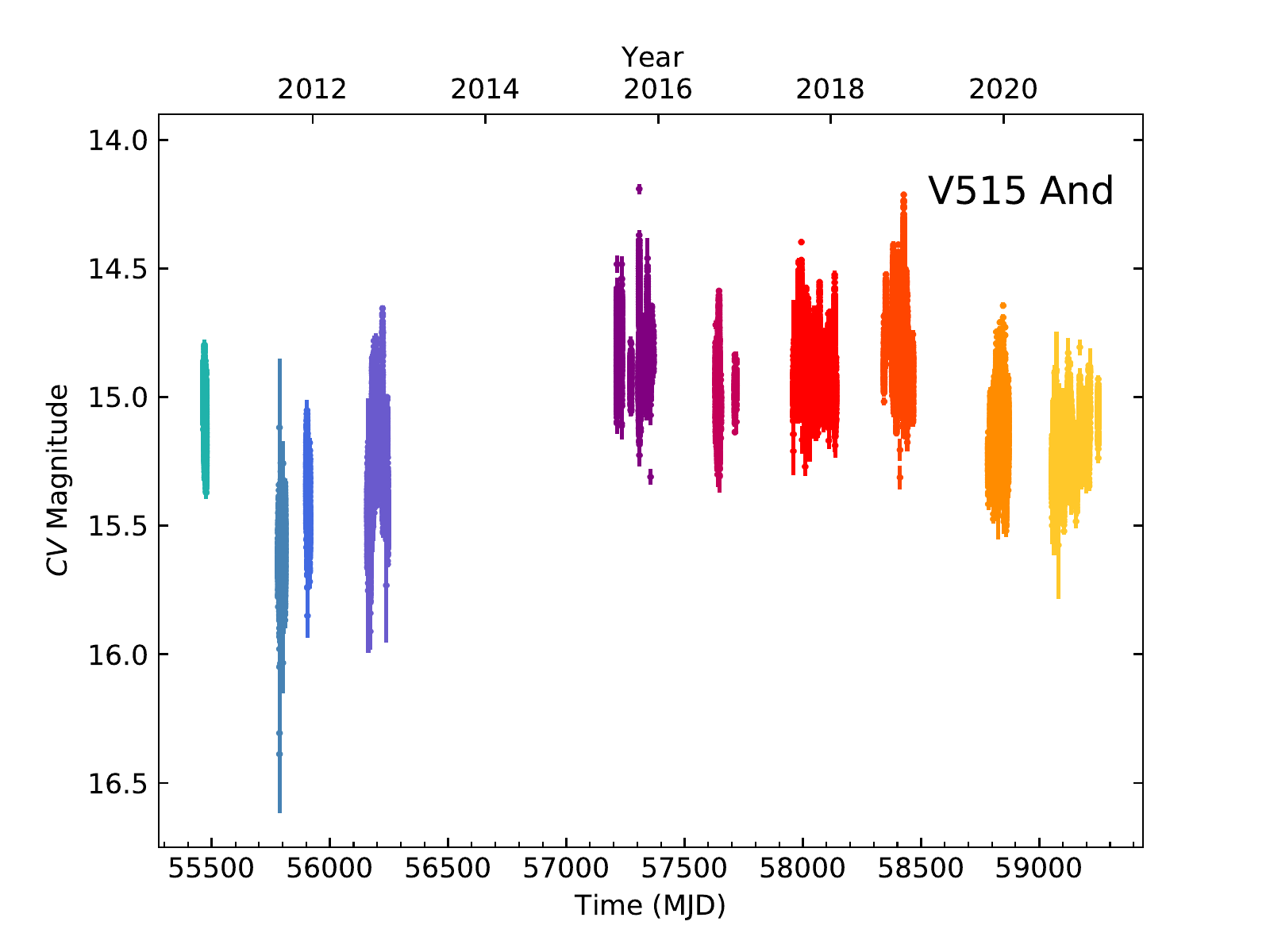}
    \caption{AAVSO light curve of V515\,And from 2011--2021. As with DW\,Cnc, observations of V515\,And were taken in a clear filter and mapped on to the $V$-band ($CV$). The colors represent the different epochs used for timing analysis.}
    \label{fig:V515And_LC}
\end{figure*}

The $CV$-band light curve of V515\,And is displayed in Fig. \ref{fig:V515And_LC}, showing $\sim$10 years of AAVSO observations. The light curve shows that the source had already begun declining when observations began, and that the source reached $CV\sim15.5$ mag by MJD $\sim55800$ before recovering to $CV\sim14.75$ by MJD $\lesssim57200$. V515\,And remained in this brighter state for $\gtrsim3$ years, before appearing to decline in flux again prior to the end of the AAVSO observations. It is unclear just from Fig. \ref{fig:V515And_LC} whether these are discrete transitions to a low state or if this is a continuous modulation in optical flux. However, we will use the terminology `low state' in this work.

\begin{figure}
    \centering
    \includegraphics[width=0.5\textwidth]{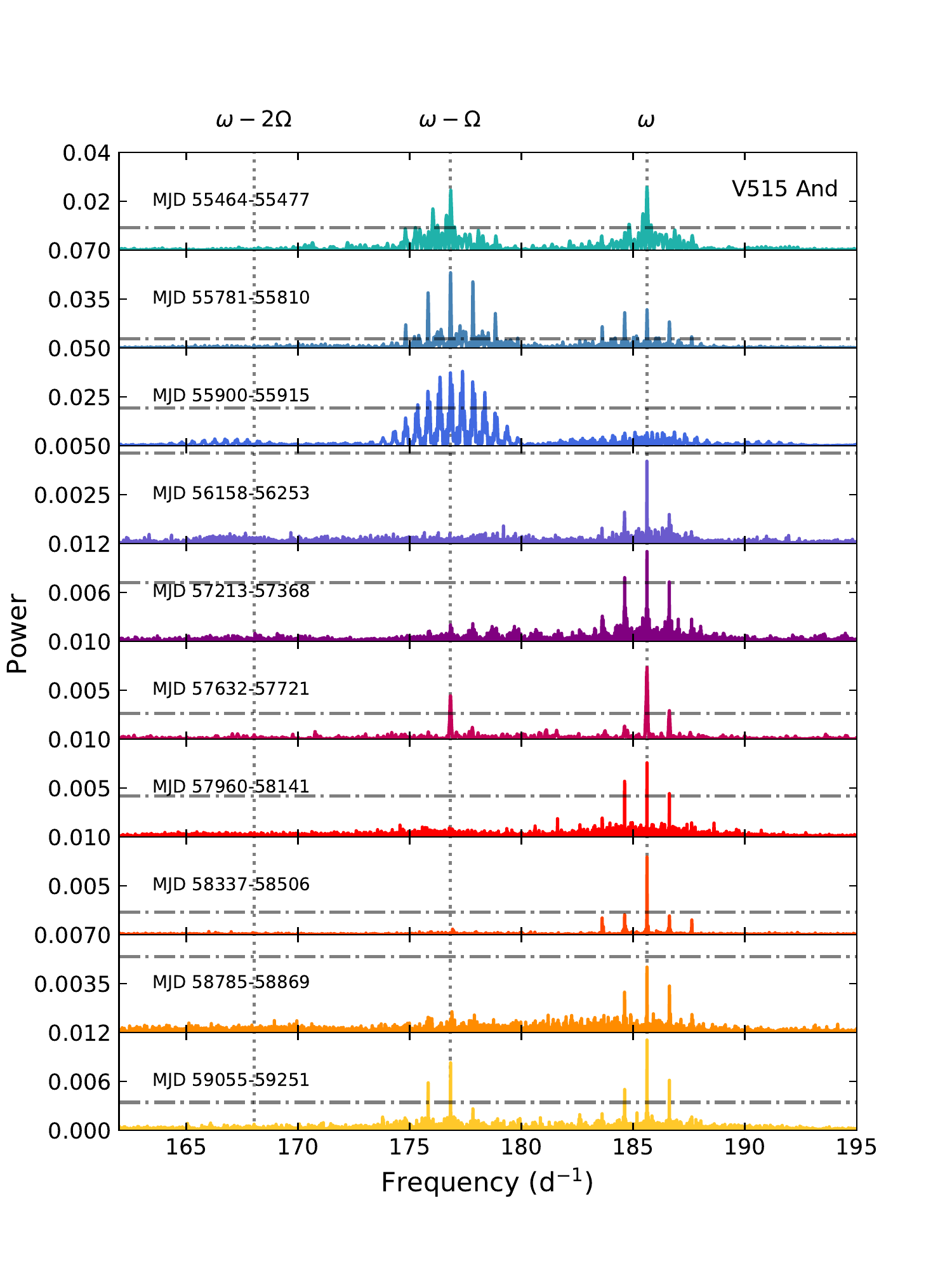}
    \caption{Lomb-Scargle periodograms of the $CV$-band light curves of V515\,And. The colors match the epochs defined in Fig. \ref{fig:V515And_LC} and the relevant MJD ranges are labeled in each panel. Vertical dotted lines and horizontal dot-dashed lines have the same meaning as in Fig. \ref{fig:DWCnc_periodograms}}
    \label{fig:V515And_periodograms}
\end{figure}

As with DW\,Cnc, we present Lomb Scargle periodograms of each epoch in Fig. \ref{fig:V515And_periodograms}. We label some characteristic frequencies associated with the system, $\omega=185.613$ d$^{-1}$ ($P_{\rm spin}=465.5$ s) and orbital sidebands $\omega-\Omega=176.825$ d$^{-1}$ and $\omega-2\Omega=168.037$ d$^{-1}$ \citep{Kozhevnikov-2012}. At the beginning of the observation period, we find the source entering a low state, and the periodograms at first show $\omega$ and $\omega-\Omega$ exhibiting equal power. As the low state progresses, the spin becomes weaker, until it is no longer significant in the third epoch, where the beat instead dominates the periodogram. The source flux then recovers, and with it so does the dominance of $\omega$ (fourth and fifth panels). A strong $\omega-\Omega$ component does appear once more in the sixth epoch, possibly associated with a slight decrease in flux, but soon disappears. In the final two epochs, we see a decline in source flux once more, though not to the levels of the first low state. Coincident with this decline we find that the beat variability starts to emerge in the periodograms once more, as seen in the final panel of Fig. \ref{fig:V515And_periodograms}.

\subsection{V1223\,Sgr}

\begin{figure*}
    \centering
    \includegraphics[width=\textwidth]{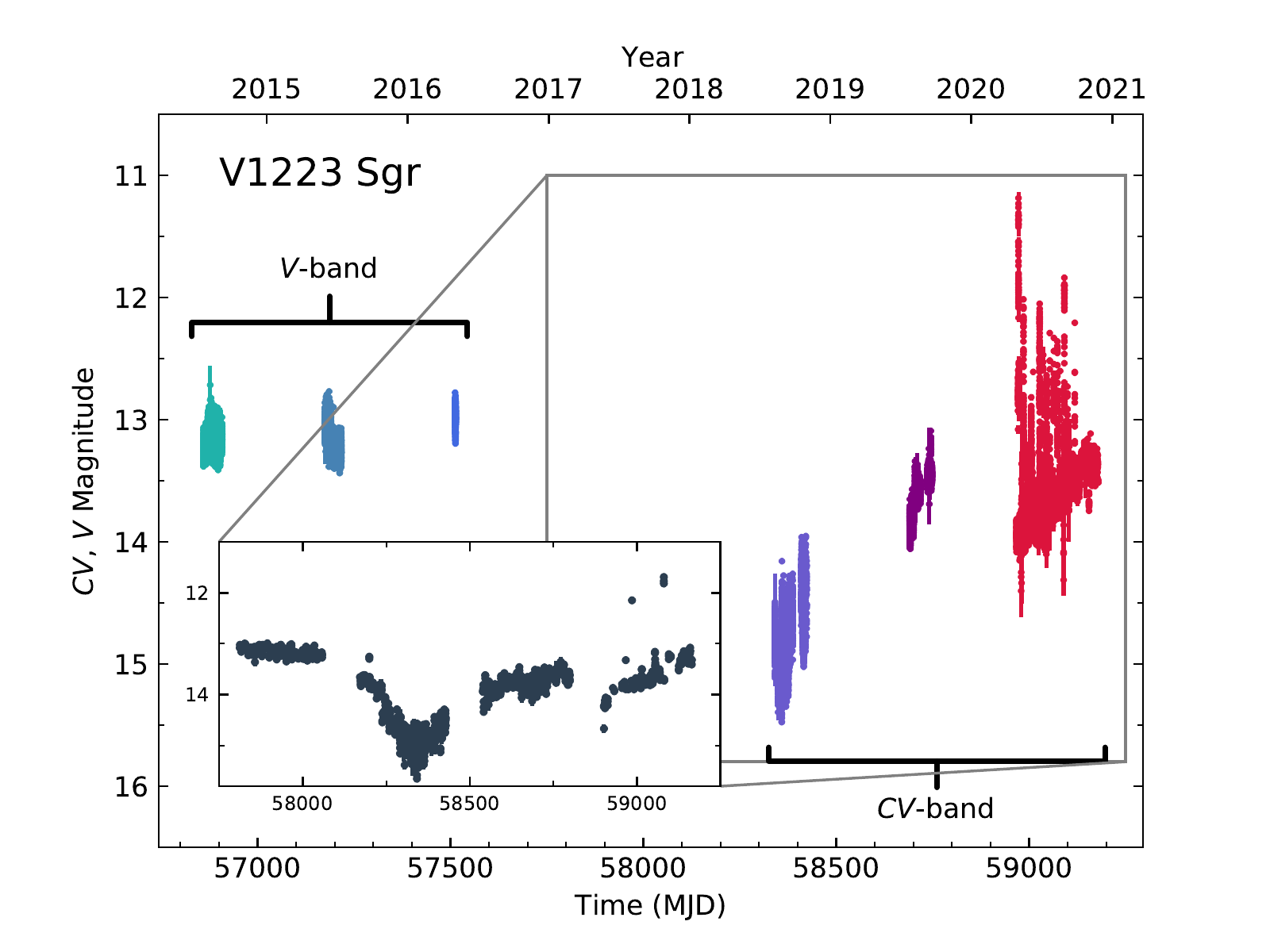}
    \caption{$CV$- and $V$-band AAVSO light curve of V1223\,Sgr from 2014--2021. The colors represent the different epochs used for timing analysis. Inset is an ASAS-SN $V$- and $g'$-band light curve from 2017--2021 which better highlights the overall shape of the light curve, including a secondary, shallower low state in late 2019.}
    \label{fig:V1223Sgr_LC}
\end{figure*}

The $CV$- and $V$-band light curve of V1223 Sgr is shown in Fig. \ref{fig:V1223Sgr_LC}, covering the observing period from 2014--2021. V1223 Sgr appears to decline in flux around or before MJD$\sim58150$ (Fig. 5; inset) before reaching its lowest flux ($CV\sim15$) around MJD$\sim58350$. The inset of Fig. \ref{fig:V1223Sgr_LC} shows that the source flux apparently recovers to near pre-low-state levels, but almost immediately declines again, reaching a local minimum at MJD$\sim58900$. This second low state is of shorter duration and lower amplitude than the first. Note in the AAVSO light curve during the recovery from the second low state, we see strong variability of up to 3 magnitudes on short ($\sim1$ d) timescales. This is reminiscent of the short duration activity that has been observed once before in the system \citep{vanAmerongen-1989,Hameury-2017a}. However, why this activity is (a) recurrent on timescales of $\sim1-2$ weeks rather than seen as a single outburst and (b) coincident with the recovery from a low state and not seen during any other epochs remains unclear. We present a possible mechanism for the outbursts in Section \ref{sec:discussion}.

\begin{figure}
    \centering
    \includegraphics[width=0.5\textwidth]{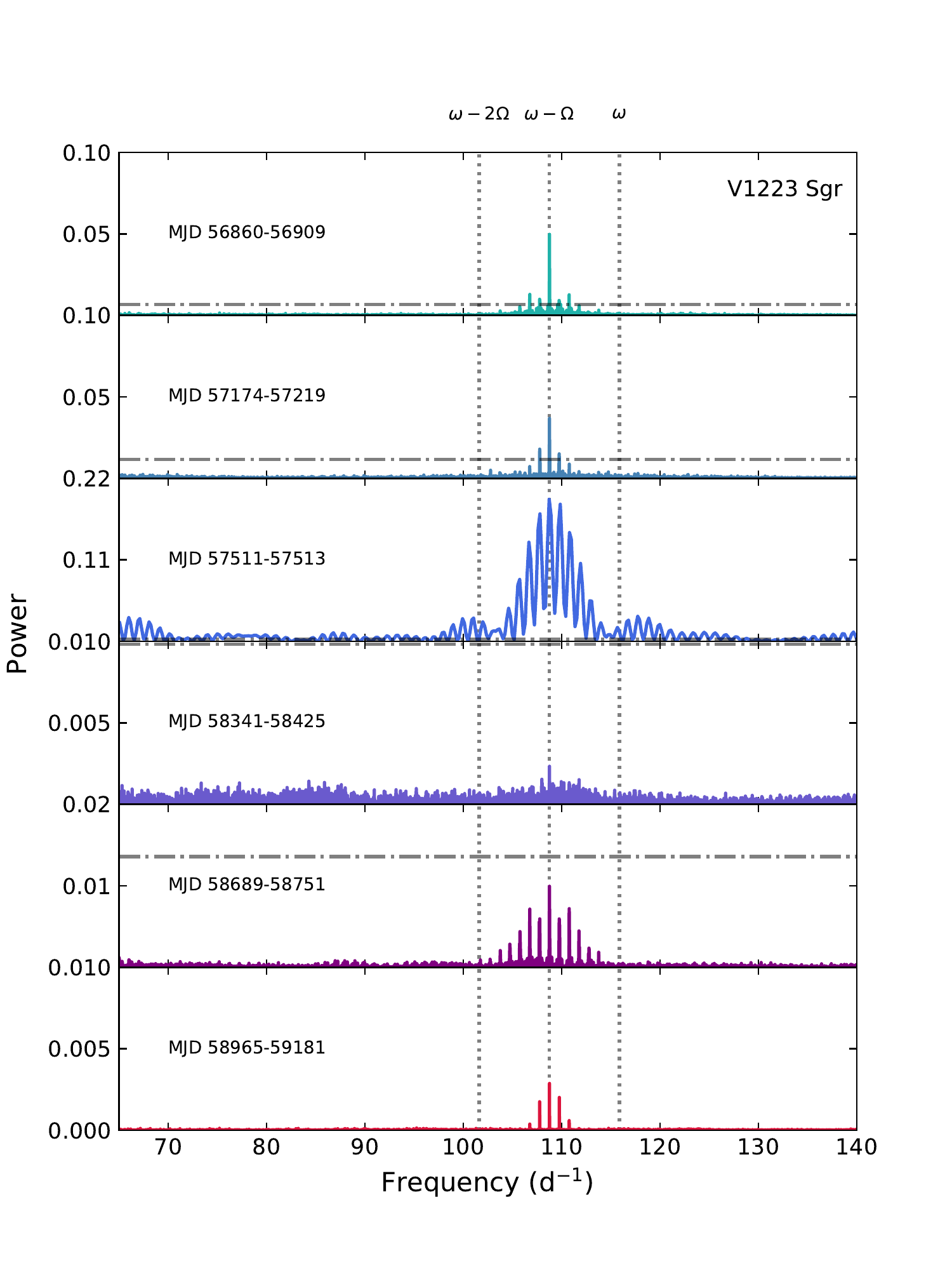}
    \caption{Lomb-Scargle periodograms of the AAVSO light curves of V1223\,Sgr. The colors match the epochs defined in Fig. \ref{fig:V1223Sgr_LC} and the relevant MJD ranges are labeled in each panel. Vertical dotted lines highlight some known periodicities associated with the system. The horizontal dot-dashed lines represent the 99.9\% confidence level. In the third panel the significance line is just visible at the bottom of the plot, in panel 4 it is at the top of the plot and in panel 6, the line lies well above the periodogram peak, beyond the limits of the $y$-axis.}
    \label{fig:V1223Sgr_periodograms}
\end{figure}

In the same manner as the other sources, we present Lomb Scargle periodograms of each epoch in Fig. \ref{fig:V1223Sgr_periodograms}. Labeled are characteristic frequencies associated with the system, $\omega$ = 115.875 d$^{-1}$ ($P_{\rm spin} = 745.63$ s) and orbital sidebands $\omega-\Omega$ = 108.744 d$^{-1}$ and $\omega-2\Omega$ = 101.615 d$^{-1}$. In the first three epochs, we see that the periodogram is dominated by the $\omega-\Omega$ signal, with some power at $\omega-2\Omega$ in the third epoch. However, in the fourth epoch, when the optical flux is at its lowest, the power at $\omega-\Omega$ all but disappears (the peak is well below the 99.9\% confidence line). The $\omega-\Omega$ power never recovers above that confidence line again - though it nearly does in the fifth epoch - despite the source flux recovering. Indeed, during the final AAVSO epoch, when the source exhibits strong variability, the signal at $\omega-\Omega$ remains weak.

\subsection{RX\,J2133.7$+$5107}

\begin{figure*}
    \centering
    \includegraphics[width=\textwidth]{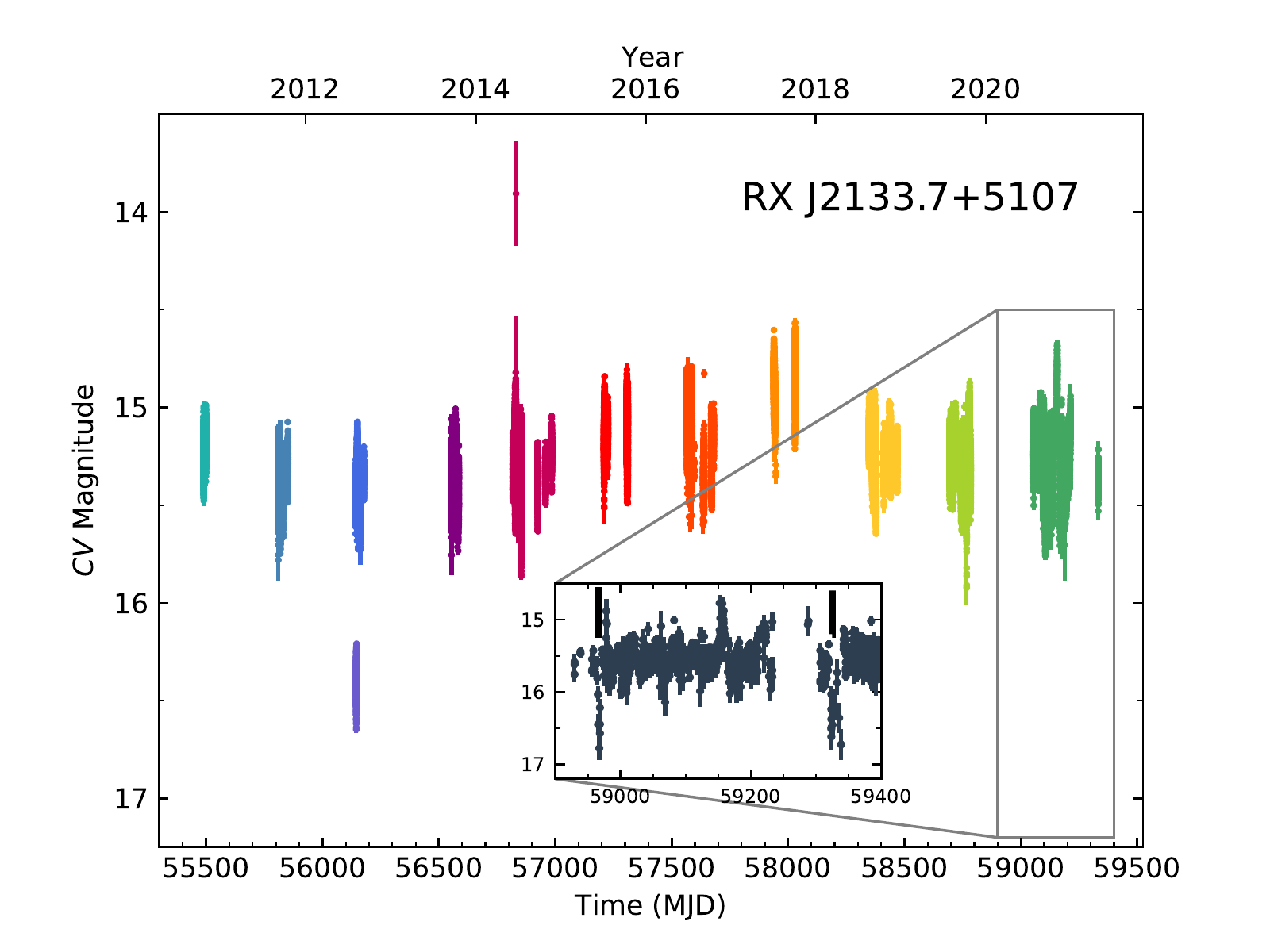}
    \caption{$CV$-band AAVSO light curve of RX\,J2133.7$+$5107 from 2010--2021. The colors represent the different epochs used for timing analysis. Inset is an ASAS-SN $g'$-band light curve from 2020--2021 highlighting two short-lived drops in flux that occurred in April 2020 and April 2021, marked by thick black lines.}
    \label{fig:RXJ2133_LC}
\end{figure*}

\begin{figure}
    \centering
    \includegraphics[width=0.5\textwidth]{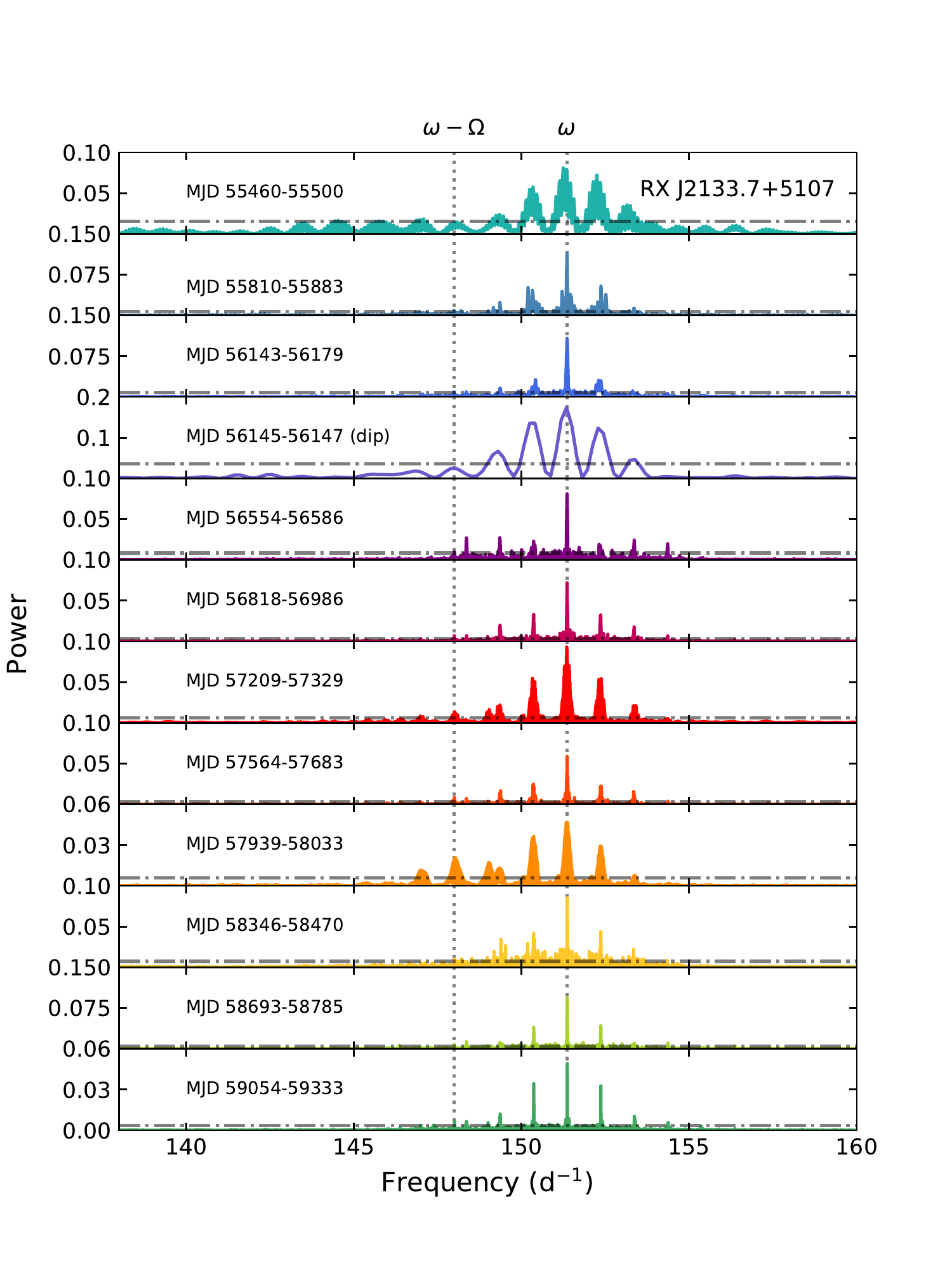}
    \caption{Lomb-Scargle periodograms of the AAVSO light curves of RX\,J2133.7$+$5107. The colors match the epochs defined in Fig. \ref{fig:RXJ2133_LC} and the relevant MJD ranges are labeled in each panel. Vertical dotted lines highlight some known periodicities associated with the system. The horizontal dot-dashed lines represent the 99.9\% confidence level.}
    \label{fig:RXJ2133_periodograms}
\end{figure}

The AAVSO $CV$-band light curve of RX\,J2133.7$+$5107 is shown in Fig. \ref{fig:RXJ2133_LC}. Over the $\sim11$ years of regular observations by the AAVSO community, the source is mostly stable, with an average $CV\sim15.2$ and showing minor sinusoidal super-orbital variability. However, the source appeared to make a brief excursion to a lower flux ($CV\sim16.5$) in August 2012 (MJD 56145--56147), before returning to its typical flux $<3$ days later. We consider the possibility that these photometric data are anomalous or calibrated incorrectly. However, we conclude that the flux decrease is real for two reasons: (a) the observer who made the putative low state measurements (B. Goff) also observed the source at later epochs, measuring magnitudes consistent with those derived by other observers during the same epochs and (b) the ASAS-SN light curve (Fig. \ref{fig:RXJ2133_LC}; inset) shows two short-lived ($\sim$days) decreases in $g$-band flux (from $g\sim15.5$ to $g\sim17$) in 2020 and 2021 (MJD$\sim58965$ and $59325$), suggesting that the August 2012 incident was not a one-off event.

The Lomb Scargle periodograms of RX\,J2133.7$+$5107 are shown in Fig. \ref{fig:RXJ2133_periodograms}. In all epochs there is strong variability at $\omega=151.361$ d$^{-1}$ ($P_{\rm spin}=570.8$ s), and we do not see this change during the short-lived low state. Interestingly, in the ninth epoch, we see that a signal at $\omega-\Omega=147.999$d$^{-1}$ emerges, but it is apparently not correlated with a drastic change in optical flux, though we note that the source is $\sim0.25$ mag brighter than the following (tenth) epoch. On closer inspection, 9 of the 12 epochs of AAVSO data show power (above the 99.9\% confidence threshold) at $\omega-\Omega$, but only in epoch 9 is it of comparable strength ($\sim50$\% power) to the spin modulation.

\subsection{Other sources showing low states}

Below we present results from two sources that exhibit definite transitions to low-flux states, but the timing analysis was inclusive either due to either a lack of periodic modulations (DO\,Dra), or not enough high-cadence (i.e. spin-resolved) photometry (V1025\,Cen). Nevertheless we present the light curves of these sources here to maintain a record of their low states. In the case of V1025\,Cen, this work represents the first published instance of the system in a low state.

\subsubsection{DO\,Dra}

\begin{figure}
    \centering
    \includegraphics[width=0.5\textwidth]{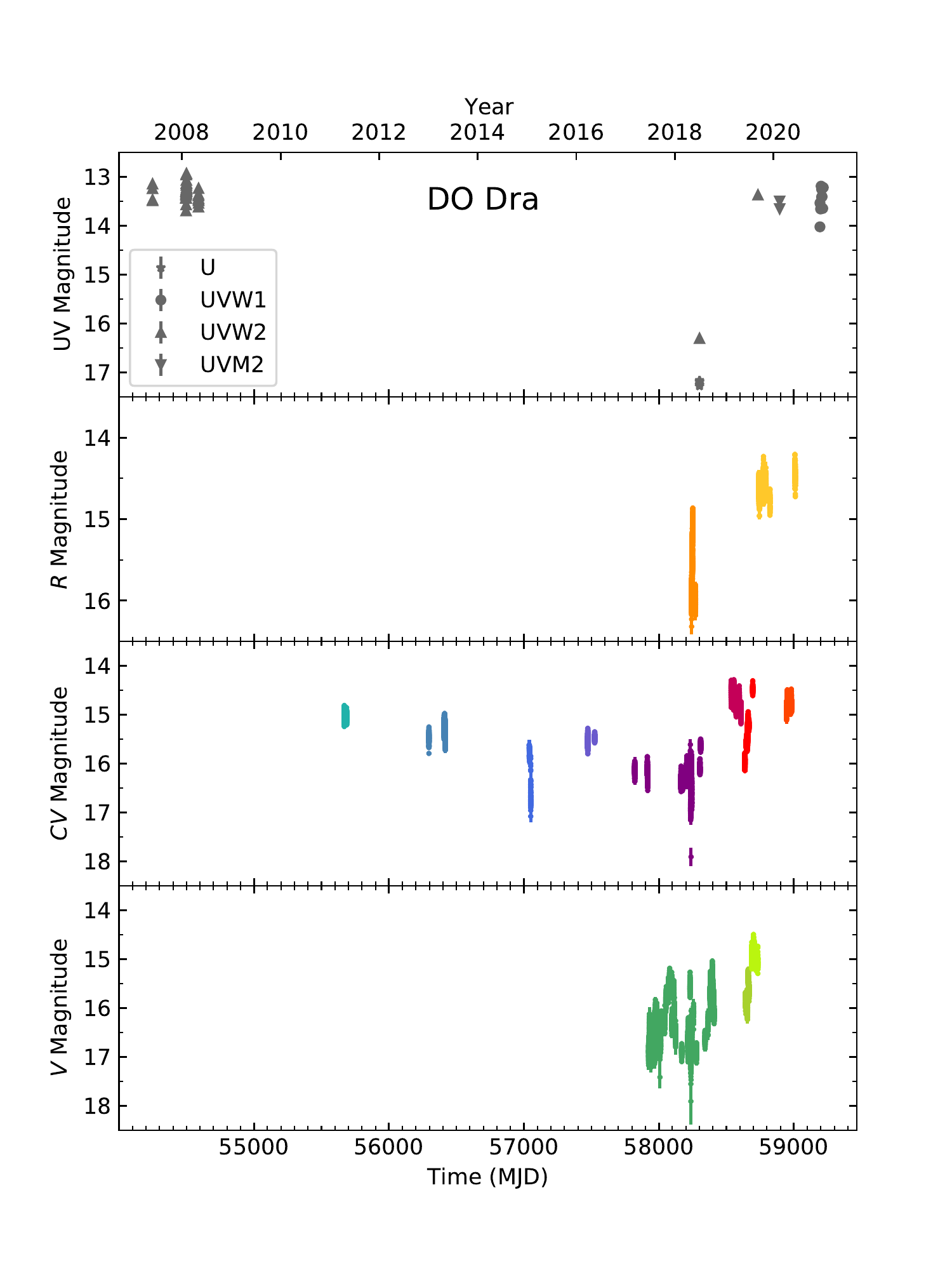}
    \caption{Multi-band light curve of DO\,Dra. The top panel shows the UV magnitudes of the source, as measured by Swift/UVOT. The remaining panels show AAVSO data in the (from top to bottom) $R$-, $CV$- and $V$-bands. The colors represent the different observing epochs used for timing analysis, though ultimately we do not find any periodicities in the periodograms so do not show them here.}
    \label{fig:DODra_LC}
\end{figure}

Multi-band light curves of DO\,Dra are shown in Fig. \ref{fig:DODra_LC}. In the top panel of Fig. \ref{fig:DODra_LC} we show the Swift/UVOT light curve in four UV filters. Swift performed just one or two exposures per visit, rather than time series observations, such that the light curve provides a better idea of the overall behavior of the average flux. As a result, we are able to see that DO\,Dra fades by $\sim3$ mag in the $UVW2$-band. The remaining panels in Fig. \ref{fig:DODra_LC} show the optical light curve in the $R$-, $CV$- and $V$-bands as observed by members of the AAVSO. The 2017--2018 low state is covered in all three bands, though we note strong ($1-2$ mag) variability, particularly in the $V$-band.

Despite the evidence of variability and sufficient high-cadence AAVSO data, a Lomb Scargle period search failed to recover any coherent signals. Though a distinct lack of periodic variability is contradictory to what has been exhibited the other sources studied in this work thus far, it is perhaps not unexpected. The spin variability in DO\,Dra has been known to be difficult to detect across all bands \citep[see e.g.][]{Patterson-1993,Haswell-1997}, and often only harmonics (usually $2\omega$) are detected \citep{Haswell-1997}. We do not detect variability at $2\omega$ or any other harmonics in the AAVSO light curves, consistent with the findings of \citet{Andronov-2018}, whose $V$-band light curves are in the AAVSO database and included in Fig. \ref{fig:DODra_LC}.  Hill et al. 2022 (submitted) also investigated the timing properties of DO Dra with the Transiting Exoplanet Survey Satellite (TESS) and find an additional, extremely short-lived ($\sim1$ d) low state in 2020.

\subsubsection{V1025\,Cen}
\label{sec:v1025}

\begin{figure}
    \centering
    \includegraphics[width=0.5\textwidth]{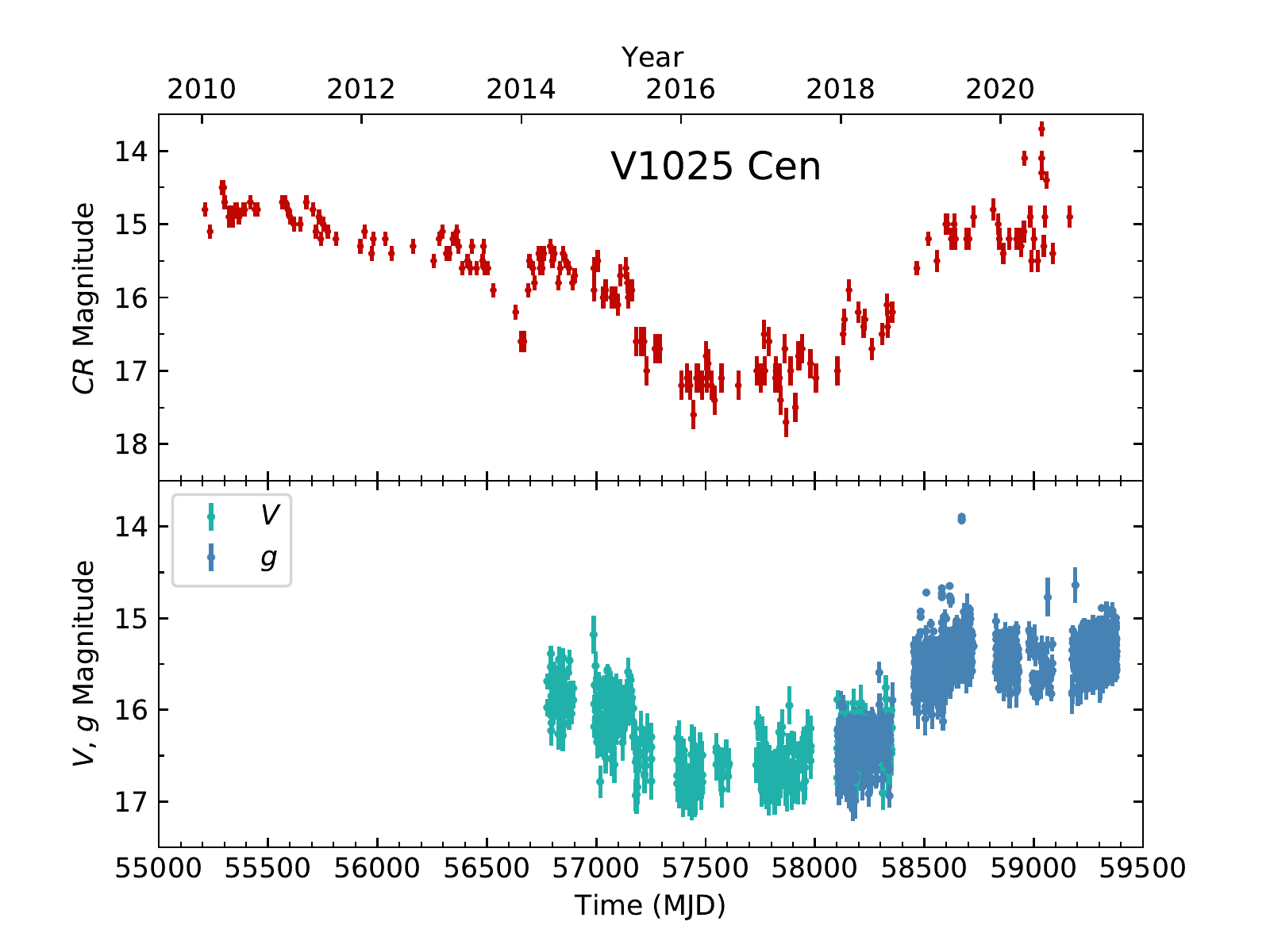}
    \caption{Multi-band light curve of V1025\,Cen. The top panel shows the AAVSO $CR$-band light curve. The bottom panel shows the $V$- and $g$-band light curves from ASAS-SN.}
    \label{fig:V1025Cen_LC}
\end{figure}

Optical light curves of V1025\,Cen are shown in Fig. \ref{fig:V1025Cen_LC}. Both ASAS-SN and AAVSO show a long-lived ($>6$ yr), high amplitude (up to $2.5$ mag) low state. This is the first time a low state has been reported for this source. Unfortunately, there are not enough high-cadence data on which to perform timing analysis, so we are unable to investigate the properties of V1025\,Cen beyond cataloging the low state.  According to the AAVSO archive, V1025\,Cen appears to have exhibited at least one more low state prior to the multi-year one shown in Fig. \ref{fig:V1025Cen_LC}, with the source reaching as faint as $CR\approx17.2$ in 2002. Additionally, the ASAS-SN light curve in the lower panel of Fig. \ref{fig:V1025Cen_LC} shows evidence of outbursting behavior as the source flux recovers. This behavior is reminiscent of what we observe in V1223\,Sgr, which also showed rapid outbursts during a recovery stage. ASAS-SN does not have the cadence required to study the outbursts in detail, such that we cannot draw accurate comparisons with V1223\,Sgr. However, \citet{Littlefield-2022} investigated the outbursts with TESS and find them to be consistent with magnetically gated accretion (see Section \ref{sec:discussion}).

\section{Discussion}
\label{sec:discussion}

We have presented optical light curves of six IPs that have shown decreases in optical flux known as low states. If we also consider FO\,Aqr, which has been extensively discussed in literature \citep[e.g.][]{Kennedy-2017,Littlefield-2020}, and AO\,Psc, which was seen to be in a low state in 1946 \citep{Garnavich-1988}, it is becoming clear that low states in IPs are not uncommon events. The light curves also show that there is diversity in the amplitude of the observed low states, with some sources exhibiting decreases in flux of $\Delta CV>2$ mag (e.g. DW\,Cnc, V1223\,Sgr) and others $\Delta CV\lesssim1$ mag  (V515\,And, RX\,J2133.7$+$5107). In addition, the duration of the low states varies greatly from source to source, with those of RX\,J2133.7$+$5107 typically lasting $<2$ weeks (Fig. \ref{fig:RXJ2133_LC}), whilst DW\,Cnc remained in a low state for over a year (Fig. \ref{fig:DWCnc_LC}).

We have also presented Lomb Scargle periodograms four of the six IPs. Timing analysis is often the key to understanding the accretion processes in IPs, and has been seen to be a useful diagnostic in the multiple low states of FO\,Aqr, both at X-ray \citep{Kennedy-2017} and optical \citep{Littlefield-2016,Littlefield-2020} wavelengths. As the flux of FO\,Aqr decreased, the optical power spectra shifted from being dominated by the spin of the WD, $\omega$, to a dominant $\omega-\Omega$ and $2(\omega-\Omega)$ component, indicative of a transition to either a disk-overflow or stream-fed accretion geometry \citep{Littlefield-2020}. 

The classic picture of an IP is one in which the WD accretes via a partial accretion disk whose inner edge terminates at the magnetospheric boundary $R_{\rm mag}$ (assuming spin equilibrium). This disk-fed geometry only occurs if the minimum distance of closest approach to the WD by the ballistic stream from $L_1$, $R_{\rm min}$ is greater than the radius of the WD $R_{\rm WD}$, otherwise no disk is able to form. Similarly, if $R_{\rm min} < R_{\rm mag}$, the ballistic stream will simply flow along the magnetic field lines on to the magnetic poles of the WD, as no disk will be able to form. \citet{King-1991} calculated that, for IPs in spin equilibrium (a reasonable assumption for most systems), then partial disks are expected to exist in systems where $P_{\rm spin} \leq 0.1 P_{\rm orb}$. If this inequality is broken, then either the system is severely out of spin equilibrium, or the accretion flow is unlike the standard Keplerian disk we typically assume when we study IPs and other CVs. The overflow of matter in an accretion disk is a long-predicted feature of mass transfer in both magnetic and non-magnetic semi-detached binaries \citep{Lubow-1989}. \citet{Lubow-1989} notes that the scale-height of the ballistic stream is typically a factor 2--3 larger than that of the disk, such that, as the gas hits the disk, it is possible for it to overflow at its outermost edge before being incorporated into the rest of the disk. However, for disk-overflow accretion {\it on to the WD itself} to occur, the ballistic stream needs to approach the inner edge of the disk, i.e. the condition $R_{\rm min} \sim R_{\rm mag}$ needs to be satisfied. As predicted by \citet{King-1991}, and noted above, this occurs in systems where $P_{\rm spin} \sim 0.1 P_{\rm orb}$. We need to consider the ratio of $P_{\rm spin}/P_{\rm orb}$ when discussing the IPs studied in this work.

\citet{Wynn-1992} presented power spectra of simulated X-ray light curves of diskless IPs, finding strong peaks at orbital sidebands $\omega-\Omega$, $2\omega-\Omega$ and harmonics. \citet{Ferrario-1999} found similar results, but also showed that at certain viewing angles one might still see power at $\omega$. In addition, \citet{Ferrario-1999} calculated theoretical power spectra of diskless IPs in the optical band, finding that $\omega$ still dominates for high inclination systems, but integer multiples of $\omega-\Omega$ dominate in low inclination IPs. In (low inclination) stream-fed IPs where the magnetic field is a perfectly symmetric dipole, \citet{Ferrario-1999} find that the power spectrum will be dominated by a peak at 2($\omega-\Omega)$. However, if this symmetry is broken in any way (i.e. material preferentially accretes on to one pole) then the light curves will be modulated on $\omega-\Omega$. Despite the many permutations of orbital inclination and inclinations of the dipole magnetic field (which affects the visibility of the hotspots), it is clear that stream-fed, disk-fed and disk-overflow IPs can have very specific timing properties. We therefore discuss the timing properties of DW\,Cnc, V515\,And, V1223\,Sgr and RX\,J2133.7$+$5107 in the context of their low states. 

\subsection{DW\,Cnc}

The four IPs of which we present Lomb Scargle periodograms all exhibit changing timing properties with time. DW\,Cnc is one of the clearest examples of this in this work. DW\,Cnc is one of the few IPs in which $P_{\rm spin} > 0.1 P_{\rm orb}$, where $P_{\rm spin} = 38.6$ min and $P_{\rm orb} = 86.1$ min \citep{Patterson-2004}. According to \citet{King-1991}, this implies that, even in the high state, the WD probably does not accrete from a standard Keplerian disk. Despite this, 
during epochs 1--3 and 6 and 7, the power spectra in Fig. \ref{fig:DWCnc_periodograms} are indicative of a typical, disk-dominated IP with power at $\omega$. Archival X-ray observations support this, showing the strongest variability at $\omega$ \citep{Nucita-2019}, typical of disk-fed IPs \citep{Ferrario-1999}. \citet{Norton-2004} find that, for IPs where $P_{\rm spin}/P_{\rm orb}\gtrsim0.5$, accretion may be fed from a ring-like structure at the edge of the WD's Roche lobe, possibly in the form of diamagnetic blobs \citep{King-1999}. DW\,Cnc falls into this category \citep{Patterson-2004,Norton-2008}, so we suggest here that the strong power seen at $\omega$ in epochs 1--3 and 6 and 7 may be due to accretion from this non-Keplerian structure. However, if, despite the apparent theoretical difficulties, a Keplerian disk did manage to form in DW\,Cnc, said disk would be a valid interpretation of the observed timing signatures in epochs 1--3, 6 and 7. Time-resolved optical photometry is not able to distinguish between the two scenarios.

In epoch 4 (MJD 58077--58221; 2017 Nov--2018 Apr) the emergence of a strong signal at $2(\omega-\Omega)$, with a slightly higher peak power than that at $\omega$, implies a change in accretion geometry. In the context of the  \citet{Ferrario-1999} model, power at $\omega-\Omega$ and/or $2(\omega-\Omega)$ in the optical power spectrum points to a stream-fed accretion flow.\footnote{This applies if the system is at a low inclination angle, which is implied by the lack of complete occulations in the X-ray light curves of DW\,Cnc \citep{Nucita-2019}} 
The presence of power at both $\omega$ and $2(\omega-\Omega)$ would typically imply a disk-overflow geometry. However, we have established previously that DW Cnc  should not be able to form a Keplerian disk, according to published theory.  
Instead, if accretion from the non-Keplerian ring-like structure of \citet{Norton-2004,Norton-2008} is the typical accretion mode of DW\,Cnc, we suggest that the WD in DW\,Cnc was accreting via both the stream and the ring-like structure in epoch 4. Again, we emphasize that, if a Keplerian disk is able to form in this system, then a disk-overflow geometry would be a valid interpretation of the data.

During epoch 5 (MJD 58409--58576; 2018 Oct--2019 Apr), the disappearance of $\omega$ from the periodogram gave way to almost equal power from $\omega-\Omega$ and $2(\omega-\Omega)$, implying a transition to a purely stream-fed accretion mode. Much like with FO\,Aqr \citep{Littlefield-2020} we find that the mode of accretion in DW\,Cnc is intimately related to the luminosity of the system, implying that mass-transfer rate ($\dot{M}$) and accretion mode are linked.

The appearance of strong variability on both $\omega-\Omega$ and $2(\omega-\Omega)$ can be explained in the \citet{Ferrario-1999} framework by an asymmetry between the upper and lower hemispheres of the magnetosphere. However, the lack of a peak at $\omega-\Omega$ in epoch 4 
is puzzling. It is possible that the stream-fed component of the accretion flow in epoch 4 is being accreted equally on to both poles, but why the stream-fed accretion flow would then become asymmetric in the following epoch is unclear. Splitting epochs 4 and 5 into shorter segments of $\sim30-60$ d each (depending on the amount of available data) and performing Lomb Scargle timing analysis reveals that there was no evolution from $2(\omega-\Omega)$ to $\omega-\Omega$ with time, i.e. epoch 4 always shows $\omega$ and $2(\omega-\Omega)$ and epoch 5 always shows $\omega-\Omega$ and $2(\omega-\Omega)$ with no change in relative power. We can only gain a complete picture of the accretion flow in these epochs with simultaneous X-ray observations, which do not exist for DW\,Cnc.

\subsection{V515\,And}

V515\,And, like DW\,Cnc, also shows clear evolution in its timing properties in Fig. \ref{fig:V515And_periodograms}. At the beginning of the AAVSO coverage of the system, it appeared to be in the midst of a transition to a low state, and the first three epochs seem to imply a transition from a disk-overflow to a stream-fed geometry as the flux declined.  
With $P_{\rm spin}/P_{\rm orb}=0.047$ \citep{Kozhevnikov-2012}, V515\,And is able to form a partial disk. The declining flux implies a shrinking of the Keplerian disk in epochs 1--2, allowing the stream to overflow, before the disk diminished completely and accretion was mostly stream-fed in epoch 3. As the flux returned to its typical high state in the fourth and fifth epochs, the WD spin re-emerged, suggesting that the disk had been replenished, though the periodogram in the fourth epoch does not quite surpass the 99.9\% confidence level line.

V515\,And showed two more epochs of likely disk-overflow accretion: in the sixth and tenth epochs, with the transition in the sixth epoch associated with only a $\sim0.1$ magnitude decrease relative to the preceding and following epochs, a decrease which is also present in the public ASAS-SN light curves. The transition in the tenth epoch is associated with a more appreciable decline in source flux of $\sim0.3-0.4$ magnitudes relative to the typical high-state flux, though still not as large a decrease as the one seen in the second and third epochs. Much like DW\,Cnc, the relative importance of the stream in V515\,And appears to be correlated with $\dot{M}$. We note that our disk-overflow hypothesis is based entirely on the periodograms of optical data, and that beat modulations at optical wavelengths can also arise from reprocessing in the binary rest frame \citep{Warner-1986}. However, we prefer the disk-overflow interpretation in V515\,And because the change from a spin-dominated to a beat-dominated regime, coincident with a decrease in flux, is more indicative of a change in accretion geometry than of a sudden onset of reprocessing in the system. We acknowledge that this interpretation isn't as definitive as if we had observed (simultaneous) beat-modulated X-rays.

\subsection{V1223\,Sgr}

V1223\,Sgr is where the interpretation becomes much more complicated. From Fig. \ref{fig:V1223Sgr_periodograms} we see that $\omega$ is never detected in the periodogram. This is not a new property of V1223\,Sgr as the orbital sidebands ($\omega-\Omega$ and occasionally $\omega-2\Omega$) are usually the dominant modulations at optical wavelengths \citep[e.g.]{Watts-1985,Warner-1986}. In the context of the \citet{Ferrario-1999} model, the lack of $\omega$ in the power spectrum, combined with the implied low inclination \citep[$i=16-40\degree$;][]{Watts-1985} might lead us to conclude that the source is a pure stream-fed system. However, the presence of a strong $\omega$ component in the X-ray light curves \citep{Osborne-1985} is compatible with a disk-fed accretion flow where the optical emission modulated on $\omega-\Omega$ is a result of spin-modulated X-rays reprocessed in the binary orbital frame, either by the hotspot or the secondary itself \citep{Warner-1986}. It is also worth noting that disk-overflow accretion is unlikely, at least in the high state, as $P_{\rm spin}/P_{\rm orb}\approx0.06$ \citep{Osborne-1985,Jablonski-1987}, such that the stream should not reach the inner-edge of the disk.\footnote{However, \citet{Patterson-2020} note that V1223\,Sgr shows prolonged episodes of spin-down, so perhaps the source is not in spin-equilibrium.} Unfortunately, without simultaneous X-ray observations during the low state, we are unable to assess whether V1223\,Sgr showed evidence of transitioning between accretion modes.

The low state and subsequent recovery of the optical flux of V1223\,Sgr was followed by an additional, smaller drop in flux, characterized by an almost complete quenching of the $\omega-\Omega$ modulation. Though the power is higher in epoch 5 than it is in epoch 4 and 6, it is still not significant (at the 99.9\% level). It is possible that the lack of the periodic signal seen in epoch 4 indicates that accretion on to the WD had ceased almost entirely and only the donor and WD were responsible for the measured optical light. To investigate this we utilized the distance to V1223\,Sgr as measured by Gaia \citep[$d=561\pm8$ pc;][]{Bailer-Jones-2021},  along with the measured extinction $A_V=0.47\pm0.02$ \citep[from $E(B-V)=0.150\pm0.005$;][]{Bonnet-Bidaud-1982}, to calculate the absolute $V$-band magnitude $M_V$ of the source at its lowest flux during the 2018 low state. We measure $M_V\sim5.8$ from the ASAS-SN $V$-band light curve, a value typical of a K0V type star \citep{Pecaut-2013}. The companion star has never been detected in the V1223\,Sgr system \citep[see e.g.][]{Watts-1985}, so we cannot confirm its spectral type. However, the measured $P_{\rm orb}=3.37$ h \citep{Jablonski-1987} implies that the donor must be of spectral type M4V with $M_V=11.6$, and the WD itself should be brighter, at $M_V=10.7$ \citep{Knigge-2011}. As $M_V\sim5.8$ is much brighter than what is expected for a CV with the orbital parameters of V1223\,Sgr, we therefore conclude that accretion was still ongoing during the deep low state in 2018, though at a lower rate than typical. The reasoning behind the weakened $\omega-\Omega$ signal during the low state is therefore unclear. One possible scenario is that the site at which reprocessing takes place \citep[i.e. the hotspot on the edge of the disk][]{Osborne-1985} is smaller at lower $\dot{M}$. Furthermore, it is also possible that there were simply fewer X-ray photons to reprocess during the low state.

During the recovery from the second flux drop, V1223\,Sgr underwent intense outbursting behavior and the large scale variations could wash out any detected periodic variability. \citet{Hameury-2017a} note that the disk instability model, which has been very successful in explaining dwarf novae (DNe) in most CVs, cannot reproduce the short duration outbursts seen in some IPs, including V1223\,Sgr. It is likely that the rapid, short-lived outbursts in V1223\,Sgr occur via another mechanism and we therefore stop short of giving them the designation of DNe here. For example, the strong magnetic field of the WD could interact with the inner disk, acting as a centrifugal barrier and preventing accretion until enough gas accumulates in the inner disk to cross the barrier and be accreted by the WD \citep{DAngelo-2010,DAngelo-2011,DAngelo-2012}. This magnetically gated accretion mechanism is thought to be responsible for the short-duration outbursts seen during the low states of MV\,Lyr \citep{Scaringi-2017} and TW\,Pic \citep{Scaringi-2021}. In addition, we note in Section \ref{sec:v1025} that the ASAS-SN light curve of V1025\,Cen (Fig. \ref{fig:V1025Cen_LC}; lower panel) also shows evidence of similar behavior during the recovery from its low state. \citet{Littlefield-2022} investigate the outbursts in V1025\,Cen and do indeed find them to be consistent with the magnetically gated accretion scenario. A detailed study of the outbursts in the recovery state of V1223\,Sgr is beyond the scope of this work, but bears further investigation through modeling the outburst profiles.

\subsection{RX\,J2133.7$+$5107}

The flux of RX\,J2133.7$+$5107 remained relatively constant throughout the AAVSO observations presented in Fig. \ref{fig:RXJ2133_LC} aside from a brief decrease in 2012. The duration of the 2012 low state is consistent with similarly short-lived ($\lesssim1$ week) flux decreases seen in ASAS-SN data (Fig. \ref{fig:RXJ2133_LC}; inset).

The WD spin dominates the power spectra of RX\,J2133.7$+$5107 at all times, implying a standard, disk-fed accretion mode \citep[$P_{\rm spin}/P_{\rm orb}\approx0.022$;][]{Bonnet-Bidaud-2006,Thorstensen-2010}, even during the short low state. Of particular interest is the ninth epoch of the AAVSO light curve, which appears to show variability on $\omega-\Omega$ at $\sim50$\% that of the power at $\omega$. This increase in power at the beat is seen to be coincident with a minor increase in optical flux, rather than the decrease in flux that was seen in DW\,Cnc and V515\,And. However, it must be noted that the majority of the AAVSO photometry does show variability on $\omega-\Omega$ (above the 99\% confidence interval). In the context of the \citet{Ferrario-1999} model, power at $\omega-\Omega$ is indicative of a not-insignificant contribution from the stream and might be suggestive of disk-overflow accretion. However, the ratio $P_{\rm spin}/P_{\rm orb}$ for RX\,J2133.7$+$5107 is $\approx0.022$, meaning that the overflow stream should not be close to the inner edge of the disk, especially during a time of enhanced mass-transfer, and disk-overflow accretion is therefore unlikely. Instead, a more likely scenario for the enhanced power at $\omega-\Omega$ in the higher $\dot{M}$ epoch 9 is increased reprocessing of the enhanced emission by parts of the system rotating in the binary frame.

\subsection{DO\,Dra and V1025\,Cen}

In the final two systems presented in this work, V1025\,Cen and DO\,Dra, we cannot draw many conclusions without detailed timing analysis. DO\,Dra showed very high-amplitude (up to $\sim2$ mag) variability even when it was ostensibly in a low state in 2018 and Lomb Scargle timing analysis revealed no coherent signals in the data. The low state of DO\,Dra is not `well-behaved' like the other sources discussed in this work, showing evidence of a varying $\dot{M}$ that can even be as high as in the `normal' high state, but also low enough for the optical flux to drop to $V\sim17$ within a matter of days (see the lower panel of Fig. \ref{fig:DODra_LC}). An apparent $V\sim17$, combined with the derived $d=195\pm1$ pc \citep{Bailer-Jones-2021} implies an absolute $M_V\sim10.5$. \citet{Mateo-1991} derive a spectral type of M$4\pm1$V, which agrees with the expected M3.5V donor as predicted by the \citet{Knigge-2011} model, given the measured $P_{\rm orb}=3.96$ h \citep{Haswell-1997} of DO\,Dra. The donor should therefore exhibit $M_V=10.9$ and the WD itself $M_V=10.3$ \citep{Knigge-2011}, very close to the observed properties of DO\,Dra. It is therefore possible that, when DO\,Dra was at its lowest flux, accretion on to the WD had ceased, and the optical flux originated from the companion star and the WD only. This is supported by the non-detections in both soft (Swift X-ray Telescope, coincident with the UVOT observations in Fig. \ref{fig:DODra_LC}) and hard \citep[NuSTAR;][]{Shaw-2020b} X-rays.  Hill et al. 2022 (submitted) present 65 d of TESS observations of DO\,Dra and find that the only periodic signal during a 28 h low state at the beginning of 2020 is the ellipsoidal modulation of the binary, further supporting the hypothesis that accretion can completely cease in this system.

Similarly, V1025\,Cen also reached as low as $V\sim17$, and has a similar distance to DO\,Dra \citep[$d=196\pm3$ pc][]{Bailer-Jones-2018}, again implying $M_V\sim10.5$. Unlike DO\,Dra, there is no published spectral type of the companion in V1025\,Cen, though the measured $P_{\rm orb}=1.41$ h \citep{Buckley-1998} implies M7.5V \citet{Knigge-2011}, which would suggest $M_V=17.8$ according to the model. However, \citet{Knigge-2011} also computes the expected $M_V$ for the WD, finding that we should expect $M_V\sim12$ for a WD in a $P_{\rm orb}=1.41$ h binary, with the exact value dependent on the effective temperature and radius of the WD. Though this is still fainter than the measured $M_V$ of V1025\,Cen during its low state, we must note that the source is likely blended in ASAS-SN \citep{Littlefield-2022}, which uses a $16\arcsec$ radius circular aperture for photometry \citep{Kochanek-2017}. A search of the Gaia Early Data Release 3 database \citep{Gaia-2016,Gaia-2021} reveals two stars within $16\arcsec$ of V1025\,Cen, one of which is only $\Delta G=0.36$ mag fainter than V1025\,Cen itself. The true magnitude of V1205\,Cen during its low state is therefore likely fainter than $V\sim17$, such that it is indeed possible that the flux was originating from only the WD and its companion, i.e. a cessation of accretion. However, we are unable to confirm this without available timing analysis  or X-ray observations during the low state.

V1025\,Cen remains a source of high interest, as it has been noted before that the timing properties do not align particularly well with stream-fed, disk-fed or disk-overflow accretion models \citep{Hellier-2002b}. The ratio $P_{\rm spin}/P_{\rm orb}=0.42$ \citep{Buckley-1998}, like DW\,Cnc, implies that the system shouldn't be able to form a Keplerian disk, unless the system is severely out of spin equilibrium. In addition, the X-ray light curve does not vary on the orbital or beat frequencies, which argues against a purely stream-fed accretion mode \citep{Hellier-1998}. However, considering that we have shown the source to exhibit multi-year low states, it is possible that early observations took place when the system was transitioning between accretion modes. \citet{Buckley-1998} presented simultaneous photometry of V1025\,Cen in the $B$- and $I$-bands, where the $B-I$ color is close to zero. Furthermore, Fig. \ref{fig:V1025Cen_LC} also suggests a Vega-like $V-CR=g-CR\sim0$ color. The photometry presented by \citet{Buckley-1998} took place in April 1995 and showed V1025\,Cen at an average $B\sim I\sim16$, lower than the typical high state (Fig. \ref{fig:V1025Cen_LC}). The spectroscopy that \citet{Hellier-2002b} used to conclude that the source does not align with any one accretion model took place in May 1996. Considering how long V1025\,Cen takes to transition between its low and high states, it is possible that the \citet{Hellier-2002b} observations did indeed occur when V1025\,Cen was in a transitional phase.

\subsection{The changing mass-transfer rate}

We have shown that for all the sources discussed in this work the source flux, and therefore $\dot{M}$, can drop to levels well below what is considered `typical' of those systems. For DW\,Cnc, V515\,And and V1223\,Sgr we can confirm that the flux decrease can be correlated with changes in their timing properties. For DW\,Cnc and V515\,And the pattern is clear: a reduction in $\dot{M}$ is coincident with a transition to a stream-fed accretion mode, via an accretion mode that is either disk-overflow (V515\,And) or possibly a hybrid between stream-fed and accretion from a non-Keplerian ring-like structure (DW\,Cnc). In V1223\,Sgr, the timing signatures do not suggest a transition between accretion modes, but we do see the periodic signal associated with the beat weaken as the flux decreases. In RX\,J2133.7$+$5107, we do not see any changes in the timing signatures associated with the low state, although we do see an increased reprocessing component coincident with a minor increase in source flux.

The physical mechanism behind the decreases in $\dot{M}$ remains unknown, though we are able to speculate. \citet{Livio-1994} proposed that starspots on the companion passing in front of the $L_1$ point are the cause of the observed drop in mass-transfer rate in VY Scl stars, and by extension this could be applied to the IPs that show similar behavior. Starspots can remain on the stellar surface for months to several years, depending on their size \citep[see e.g.][]{Hall-1994,Hatzes-1995}, which can explain the multi-year low states observed in e.g. DW\,Cnc and V1223\,Sgr. 

Another possibility for the reduction in $\dot{M}$ could be variations in the radius of the secondary. \citet{Bianchini-1990} noted that the late-type donors in CVs can exhibit solar-type cycles, resulting in periodic variations of the quiescent luminosity of the binary. The author attributed this to fractional variations in stellar radius in the range $\Delta R/R=0.6-3.0\times10^{-4}$. Any change in the radius of the donor would cause a variation in $\dot{M}$ owing to the Roche lobe overflow nature of accretion in CVs. 

It is not clear which, if either, of the two above scenarios are responsible for the low states we have presented in this work. However, both are linked to a Sun-like cycle of activity in the donor star. In future observations of low states, we may be able to test for variations in $P_{\rm orb}$ using optical spectroscopy, which would be indicative of a change in the donor radius \citep{Bianchini-1990}, but it is possible that an entirely different mechanism is responsible for the observed behavior of these IPs.

\subsection{A census of IP low states}
\label{sec:census}

 Beside the six sources presented in this paper, there are numerous other IPs or candidate IPs that appear to exhibit low states. In this section we attempt to present an exhaustive list of all candidate and confirmed IPs that have historically shown this behavior.

{\it Confirmed IPs:} AO\,Psc \citep{Garnavich-1988}, FO\,Aqr \citep{Garnavich-1988,Kennedy-2017,Littlefield-2020}, V1223 Sgr \citep[][and this work]{Garnavich-1988}, DW\,Cnc \citep[][and this work]{SeguraMontero-2020}, V515\,And (this work), RX\,J2133.7$+$5107 (this work), DO\,Dra \citep[][Hill et al. 2022, submitted, and this work]{Andronov-2018}, V1025\,Cen \citep[][and this work]{Littlefield-2022}, V1323\,Her \citep{Andronov-2014}, 1RXS\,J211336.1$+$542226 \citep{Halpern-2018}, V1062\,Tau \citep{Lipkin-2004}.

{\it Candidate IPs:} Swift\,J0746.2$-$1611 \citep{Bernardini-2019}, J183221.56$-$162724.25 \citep{Beuermann-2021}.

\section{Conclusions}

We have presented optical observations of 6 intermediate polars that have shown decreases in flux known as low-states, two of which, V515\,And, RX\,J2133.7$+$5107 have not previously been reported to exhibit such behavior. We summarize our findings below:

\begin{itemize}
    \item Timing analysis of DW\,Cnc suggests that the source transitions to a completely stream-fed geometry at its lowest flux, before recovering its original timing properties with the recovery of the flux. According to theoretical interpretations of the $P_{\rm spin}/P_{\rm orb}$ ratio, DW\,Cnc should not be a disk-fed system in its normal, high state. We thus hypothesise that it could accrete from a ring-like structure, which dissipates as the flux decreases.
    \item The temporal properties of V515\,And suggest that the two decreases in optical flux that we see are related to a transition between disk-fed, stream-fed and disk-overflow accretion geometries.
    \item The timing signatures in V1223\,Sgr almost completely disappeared as the source entered its first of the two low states presented in this work. We investigate whether accretion on to the WD had almost entirely stopped in this phase but conclude that it was still ongoing, though at a lower $\dot{M}$ than typical for the system. We find no evidence for a changing accretion mode in V1223\,Sgr. However, the source did show rapid outbursts during its recovery from the second, shallower low state, which we suggest could be due to episodes of magnetically gated accretion.
    \item RX\,J2133.7$+$5107 is unusual in that the low states only lasted a matter of days, and the only change in the timing signatures came when the optical flux was slightly higher than typical, during which time we conclude that the system's increased $\dot{M}$ led to increased reprocessing in the binary orbital frame.
    \item DO\,Dra shows high-amplitude variability within its low state and we see no timing signatures in the Lomb Scargle periodograms. We hypothesise that, at the source's lowest flux ($M_{\rm V}\sim10.5$), accretion on to the WD had stopped.
    \item Similarly, V1025\,Cen reached an absolute $M_V\sim10.5$ at its lowest flux, possibly indicating that only the WD (and companion star) were visible and accretion had ceased, though we are unable to confirm this without available timing analysis or X-ray observations.
\end{itemize}

There are a variety of low state behaviors presented in this work, and in future we must combine optical spectroscopy, photometry and X-ray observations to gain a complete understanding of the physical mechanisms at work.

\section*{Acknowledgments}

The authors thank the anonymous referee for comments that helped improve the manuscript. AEC acknowledges partial support from the Nevada Undergraduate Research Award. AWS thanks Jean-Marie Hameury for fruitful discussions on the outbursts in V1223\,Sgr. We acknowledge with thanks the variable star observations from the {\it AAVSO International Database} contributed by observers worldwide and used in this research. AWS thanks Elizabeth Waagen of the AAVSO for facilitating the inclusion of AAVSO observers as co-authors.

\bibliography{Dipping_IPs.ACCEPTED.arxiv.bib}
\bibliographystyle{aasjournal}

\end{document}